## *SUZAKU* OBSERVATIONS OF HEAVILY OBSCURED (COMPTON-THICK) ACTIVE GALACTIC NUCLEI SELECTED BY *SWIFT*/BAT HARD X-RAY SURVEY

Atsushi Tanimoto[1], Yoshihiro Ueda[1], Taiki Kawamuro[1,2], Claudio Ricci[3,4,5], Hisamitsu Awaki[6], Yuichi Terashima[6]
[1]Department of Astronomy, Kyoto University, Kyoto 606-8502, Japan
[2]National Astronomical Observatory of Japan, Osawa, Mitaka, Tokyo 181-8588, Japan
[3]Instituto de Astrofísica, Pontificia Universidad Católica de Chile, Casilla 306, Santiago 22, Chile
[4]Kavli Institute for Astronomy and Astrophysics, Peking University, Beijing 100871, China
[5]Chinese Academy of Sciences South America Center for Astronomy and China-Chile Joint Center for Astronomy, Camino El Observatorio 1515, Las Condes, Santiago, Chile
[6]Department of Physics, Ehime University, Matsuyama 790-8577, Japan
(Received receipt date; Revised revision date; Accepted acceptance date)
*Draft version January 5, 2018*

### ABSTRACT

We present a uniform broadband X-ray (0.5–100.0 keV) spectral analysis of 12 *Swift*/Burst Alert Telescope (BAT) selected Compton-thick ($\log N_{\rm H}/{\rm cm}^{-2} \geq 24$) Active Galactic Nuclei (CTAGNs) observed with *Suzaku*. The *Suzaku* data of 3 objects are published here for the first time. We fit the *Suzaku* and *Swift* spectra with models utilizing an analytic reflection code and those utilizing the Monte Carlo based model from an AGN torus by Ikeda et al. (2009). The main results are as follows. (1) The estimated intrinsic luminosity of a CTAGN strongly depends on the model; applying Compton scattering to the transmitted component in an analytic model may largely overestimates the intrinsic luminosity at large column densities. (2) Unabsorbed reflection components are commonly observed, suggesting that the tori are clumpy. (3) Most of CTAGNs show small scattering fractions ($< 0.5$ %) implying a buried AGN nature. (4) Comparison with the results obtained for Compton-thin AGNs (Kawamuro et al. 2016a) suggests that the properties of these CTAGNs can be understood as a smooth extension from Compton-thin AGNs with heavier obscuration; we find no evidence that the bulk of the population of hard X-ray selected CTAGN is different from less obscured objects.

*Subject headings:* galaxies: active – galaxies: Seyfert – X-rays: galaxies

### 1. INTRODUCTION

To reveal the nature of heavily obscured Active Galactic Nuclei (AGNs) whose line-of-sight hydrogen column density is $\log N_{\rm H}/{\rm cm}^{-2} \geq 24$, so-called Compton-thick AGNs (CTAGNs), is an important, yet unresolved issue in modern astronomy (Ueda 2015). CTAGNs are thought to be key objects to understand the origin of the co-evolution of Supermassive Black Holes (SMBHs) and their host galaxies (Kormendy & Ho 2013). According to a galaxy/SMBH evolutionary scenario (e.g., Hopkins et al. 2006), major mergers trigger violent star formation and rapid growth of SMBHs heavily obscured by gas and dust (Ricci et al. 2017). This leads to the idea that some CTAGNs may be distinct populations (i.e., those in a different evolutionary stage) from less obscured AGNs. Due to observational difficulties in detecting CTAGNs, however, it remains an open question whether CTAGNs are intrinsically same objects or not as the rest of AGNs in terms of their nucleus structure, host galaxy properties and cosmological evolution.

The presence of CTAGN populations is required to explain the origin of the Cosmic X-ray Background (CXB) (e.g., Ueda et al. 2003; Gilli et al. 2007; Treister et al. 2009; Ueda et al. 2014; Aird et al. 2015). Their estimated number density is highly uncertain, however, depending on several parameters assumed in population synthesis models of the CXB (e.g., Akylas et al. 2012). Among them, modeling of broadband X-ray spectra of CTAGNs is a critical issue. The spectral modeling also largely affects an estimate of the intrinsic luminosity of distant CTAGNs detected in deep surveys with limited photon statistics, and hence the determination of their luminosity function. Thus, it is very important to systematically analyze high-quality broadband (0.5–100 keV) X-ray spectra of a large sample of bright CTAGNs in the local universe and thereby determine averaged spectra of CTAGNs as a function of column density.

Hard X-ray observations above 10 keV provide one of the least-biased AGN samples thanks to the strong penetrating power against obscuration unless the column density largely exceeds $\log N_{\rm H}/{\rm cm}^{-2} \geq 24.5$. All-sky hard X-ray surveys performed with *Swift* and *INTEGRAL* have produced catalogs of local AGNs including CTAGNs (Markwardt et al. 2005; Beckmann et al. 2006, 2009; Tueller et al. 2008, 2010; Burlon et al. 2011; Ajello et al. 2012; Baumgartner et al. 2013; Malizia et al. 2016), which have been extensively followed-up with pointed observations (e.g., Ajello et al. 2008; Winter et al. 2009a; Vasudevan et al. 2013). From the *Swift*/BAT 70-month catalog (Baumgartner et al. 2013), Ricci et al. (submitted to ApJS) systematically analyze X-ray data below 10 keV obtained mainly with *Swift*/X-Ray Telescope (XRT) and *XMM-Newton*, which are combined with the *Swift*/BAT spectra in the 14–150 keV. As a result, Ricci et al. (2015) identify 55 CTAGN candidates (see also Akylas et al. 2016). However, since the photon statistics of the *Swift*/XRT spectrum is limited and the data above and below 10 keV were not simultaneously obtained (thus being affected by time variability), large uncertainties remain in the derived column density and intrinsic luminosity.

The *Suzaku* observatory (Mitsuda et al. 2007) has a unique capability of simultaneously observing broadband X-ray spectra covering the 0.5–40.0 keV with high-quality charge coupled device (CCD) data below 10 keV (Section 2). The combination of *Suzaku* and *Swift* is proved to be very powerful for studying the broadband spectra of local AGNs (Ueda et al. 2007; Eguchi et al. 2009, 2011; Winter et al. 2009b; Tazaki et al. 2011, 2013; Gandhi et al. 2013, 2015; Kawamuro et al. 2013, 2016a,b). Recently, *NuSTAR* have started to observe nearby CTAGNs (e.g., Puccetti et al. 2014; Koss et al. 2015; Rivers et al. 2015; Guainazzi et al. 2016) covering the 4–80 keV. However, most of them do not have simultaneous CCD spectra in the 0.5–10 keV, which are particularly useful to observe emission line and absorption edge features thanks to the good energy resolution.

This paper presents a summary of uniform spectral analyses of local CTAGNs observed with *Suzaku* and *Swift*, following Kawamuro et al. (2016a) for Compton-thin ($22 \leq \log N_{\rm H}/{\rm cm}^{-2} < 24$) AGNs. Section 2 describes the sample selection and data reduction of the *Suzaku* data. Our sample consists of 12 CTAGNs selected from the *Swift*/BAT 70-month catalog. The *Suzaku* spectra of 3 objects are reported here for the first time. Section 3 presents the *Suzaku* light curves and the analysis of the *Suzaku*+*Swift*/BAT spectra. We apply analytical models and Monte-Carlo based torus models by Ikeda et al. (2009) to these spectra. Section 4 summarizes our results for the individual objects in comparison with earlier works. In Section 5, we discuss our overall results by comparing the Compton-thick population with the Compton-thin AGNs. The luminosities are calculated from the observed redshift with the cosmological parameters $(H_0, \Omega_{\rm m}, \Omega_\lambda) = (70~{\rm km~s}^{-1}~{\rm Mpc}^{-1}, 0.3, 0.7)$. The solar abundances by Anders & Grevesse (1989) are assumed in all cases. In modeling photoelectric absorption, we adopt the cross section given by Verner & Ferland (1996) and Verner et al. (1996). The errors on the spectral parameters correspond to the 90% confidence limits for a single parameter.

## 2. SAMPLE AND ANALYSIS

### 2.1. *Sample*

Our sample consists of 12 CTAGN candidates from Ricci et al. (2015) (i.e., a subsample of the *Swift*/BAT 70-month catalog, Baumgartner et al. 2013) that were observed with *Suzaku* during its lifetime. To ensure high spectral quality, we select data with a net exposure longer than 20 ksec. We do not include NGC 1106, NGC 2788A, UGC 03752 (Tanimoto et al. 2016), which were observed without the Hard X-ray Detector (HXD) sensitive to energies above 10 keV. We also exclude objects whose *Suzaku* data have already been thoroughly analyzed in individual papers, such as Circinus Galaxy (Yang et al. 2009), NGC 1068 (Bauer et al. 2015) and NGC 3079 (Konami et al. 2012), and two low luminosity AGNs, NGC 4102 and NGC 5643 (Kawamuro et al. 2016b). In particular, Circinus galaxy and NGC 1068 are heavily Compton-thick AGNs with line-of-sight column densities larger than $10^{25}$ cm$^{-2}$, which could be distinct populations to mildly Compton thick ones ($\log N_{\rm H}/{\rm cm}^{-2} < 25$) studied in this paper. Tables 1 and 2 show the information on the sample and the *Suzaku* observation log, respectively. The results based on *Suzaku* observations of three objects (NGC 1194, NGC 6552 and NGC 7130) are reported here for the first time. Although NGC 7582 is a changing look AGN (Bianchi et al. 2009), we include this object in the sample, which was classified as a CTAGN by Ricci et al. (2015) with the spectral analysis of the long-term averaged *Swift*/BAT spectrum and *XMM-Newton* spectra.

### 2.2. *Analysis*

*Suzaku* (Mitsuda et al. 2007) is the fifth Japanese X-ray astronomy satellite, which operated from 2005 to 2015. It carried on board four X-ray CCD cameras called the X-ray Imaging Spectrometers (XIS-0, XIS-1, XIS-2 and XIS-3) as the focal plane instruments of four X-Ray Telescopes (XRTs), and a non-imaging, collimated Hard X-ray Detector (HXD). The XIS cameras covered the 0.3–12.0 keV as focal plane detectors of the X-ray Telescopes; XIS-0, XIS-2 and XIS-3 were Front-Side Illuminated CCDs (FIXISs) and XIS-1 was the Back Illuminated one (BIXIS). Since 2006 November, XIS-2 became unusable most probably due to a micro-meteoroid impact. The HXD consisted of Si PIN photo-diodes and Gadolinium Silicon Oxide scintillation counters, which covered the 10–70 keV and 40–600 keV, respectively. We analyze the XIS and HXD data using HEAsoft version 6.21 with the calibration database (CALDB) released on 2016 February 15 for the XIS and the CALDB released on 2011 September 13 for the HXD.

#### 2.2.1. *Suzaku/XIS*

We reprocess the unfiltered XIS data by using aepipeline. To extract the light curves and spectra, we use circular regions centered on the source peak with a radius of 90 arcsec, while the background is taken from a homocentric annular region with inner and outer radii of 120 arcsec and 240 arcsec, respectively. Since the background spectra also contain about 10% of the source photons due to the tail of the point spread function of the XRT, we correct for this effect in our spectral analysis. We generate the XIS response matrix with xisrmfgen and ancillary response files with xissimarfgen (Ishisaki et al. 2007). We bin each of the XIS spectra to contain at least 50 counts per bin.

#### 2.2.2. *Suzaku/HXD*

In this paper, we only analyze the HXD-PIN data, because all of our targets except NGC 4945 were too faint at energies above 50 keV to be detected by HXD-GSO. We reprocess the unfiltered HXD data by using aepipeline. We utilize the "tuned" background event files (Fukazawa et al. 2009) to reproduce the spectrum of the Non X-ray Background (NXB), except for ESO 565–G019[1], to which the simulated spectrum of the CXB is added. In the spectral analysis, we only utilize an energy range where the source flux is brighter than 3% of the NXB level (Fukazawa et al. 2009).

---

[1] We utilize the same background spectrum as used in Gandhi et al. (2013), which was produced from night Earth data, because the reproductivity of the tuned background file was found to be insufficient.

TABLE 1
Information on Targets

| (1) Galaxy Name | (2) *Swift* ID | (3) RA | (4) DEC | (5) Redshift | (6) $N_{\rm H}^{\rm Gal}$ | (7) $\log M_{\rm BH}/M_\odot$ | (8) $M_{\rm BH}$ Ref. |
|---|---|---|---|---|---|---|---|
| CGCG 420–015 | J0453.4+0404 | 04h53m25s | +04d03m42s | 0.0294 | 0.0654 | 8.31 | (1) |
| ESO 137–G034 | J1635.0–5804 | 16h35m14s | –58d04m48s | 0.0090 | 0.2250 | 8.02 | (2) |
| ESO 323–G032 | J1253.5–4137 | 12h53m20s | –41d38m08s | 0.0160 | 0.0844 | 7.56 | (1) |
| ESO 565–G019 | J0934.7–2156 | 09h34m44s | –21d55m40s | 0.0163 | 0.0416 | ⋯ | ⋯ |
| Mrk 3 | J0615.8+7101 | 06h15m36s | +71d02m15s | 0.0135 | 0.0998 | 7.96 | (2) |
| NGC 1194 | J0304.1–0108 | 03h03m49s | –01d06m13s | 0.0136 | 0.0597 | 8.12 | (1) |
| NGC 3393 | J1048.4–2511 | 10h48m23s | –25d09m43s | 0.0125 | 0.0605 | 7.20 | (3) |
| NGC 4945 | J1305.4–4928 | 13h05m28s | –49d28m06s | 0.0019 | 0.1350 | 6.14 | (3) |
| NGC 5728 | J1442.5–1715 | 14h42m24s | –17d15m11s | 0.0093 | 0.0774 | 8.07 | (1) |
| NGC 6552 | J1800.3+6637 | 18h00m07s | +55d36m54s | 0.0265 | 0.0381 | ⋯ | ⋯ |
| NGC 7130 | J2148.3–3454 | 21h48m20s | –34d57m04s | 0.0162 | 0.0185 | 7.61 | (1) |
| NGC 7582 | J2318.4–4223 | 23h18m24s | –42d22m14s | 0.0052 | 0.0121 | 7.56 | (4) |

Note. (1) Galaxy name. (2) *Swift* ID. (3)-(5) Position and Redshift taken from NASA/IPAC Extragalactic Database (NED). (6) Galactic absorption (Kalberla et al. 2005) in units of $10^{22}$ cm$^{-2}$. (7) Logarithmic black hole mass. (8) Reference of the black hole mass.
Refe. (1) Koss et al. (2017). (2) Khorunzhev et al. (2012). (3) van den Bosch (2016). (4) Izumi et al. (2016).

TABLE 2
*Suzaku* Observation Log

| (1) Galaxy Name | (2) *Suzaku* ID | (3) Start Date | (4) End Date | (5) Exposure | (6) Nominal Position | (7) *Suzaku* Ref. |
|---|---|---|---|---|---|---|
| CGCG 420–015 | 704058010 | 2009-09-01 | 2009-09-04 | 109.1 | HXD | (1) |
| ESO 137–G034 | 403075010 | 2008-10-05 | 2008-10-07 | 92.1 | HXD | (2) |
| ESO 323–G032 | 702119010 | 2007-12-22 | 2007-12-24 | 79.2 | HXD | (2) |
| ESO 565-G019 | 707013010 | 2012-05-20 | 2012-05-22 | 78.9 | XIS | (3) |
| Mrk 3 | 100040010 | 2005-10-22 | 2005-10-24 | 95.0 | HXD | (4) |
| NGC 1194 | 704046010 | 2009-08-01 | 2009-08-02 | 50.3 | XIS | ⋯ |
| NGC 3393 | 702004010 | 2007-05-23 | 2007-05-25 | 55.2 | HXD | ⋯ |
| NGC 4945 | 100008030 | 2006-01-15 | 2006-01-17 | 95.1 | HXD | (5) |
| NGC 5728 | 701079010 | 2006-07-19 | 2006-07-20 | 41.3 | HXD | (2) |
| NGC 6552 | 708014010 | 2013-11-14 | 2013-11-17 | 105.8 | XIS | ⋯ |
| NGC 7130 | 703012010 | 2008-05-11 | 2008-05-12 | 44.5 | HXD | ⋯ |
| NGC 7582 | 702052040 | 2007-11-16 | 2007-11-16 | 31.9 | HXD | ⋯ |

Note. (1) Galaxy name. (2) *Suzaku* observation ID. (3) Start date in units of ymd. (4) End date in units of ymd. (5) Exposure in units of ksec. (6) XIS nominal or HXD nominal. (7) Reference of the previous work using the *Suzaku* data.
Refe. (1) Severgnini et al. (2011). (2) Comastri et al. (2010). (3) Gandhi et al. (2013). (4) Awaki et al. (2008). (5) Itoh et al. (2008).

TABLE 3
Results of Time Variability Test

| (1) Galaxy Name | (2) $\chi^2$/dof (XIS) | (3) p-value (XIS) | (4) $\chi^2$/dof (HXD) | (5) p-value (HXD) |
|---|---|---|---|---|
| CGCG 420–015 | 14.2/47.0 | 1.00 | 95.0/47.0 | $< 0.01$ |
| ESO 137–G034 | 15.6/34.0 | 1.00 | 51.8/30.0 | 0.03 |
| ESO 323–G032 | 12.1/24.0 | 0.98 | 22.4/24.0 | 0.56 |
| ESO 565–G019 | 13.3/29.0 | 0.99 | 31.8/30.0 | 0.38 |
| Mrk 3 | 8.0/33.0 | 1.00 | 49.7/32.0 | 0.02 |
| NGC 1194 | 7.4/17.0 | 0.98 | 24.9/17.0 | 0.10 |
| NGC 3393 | 4.9/21.0 | 1.00 | 23.7/21.0 | 0.31 |
| NGC 4945 | 13.0/40.0 | 1.00 | 561.0/39.0 | $< 0.01$ |
| NGC 5728 | 5.3/14.0 | 0.98 | 19.3/13.0 | 0.11 |
| NGC 6552 | 11.7/33.0 | 1.00 | 193.0/33.0 | $< 0.01$ |
| NGC 7130 | 6.1/16.0 | 0.99 | 32.0/16.0 | 0.02 |
| NGC 7582 | 9.5/13.0 | 0.73 | 23.0/13.0 | 0.04 |

Note. (1) Galaxy name. (2) $\chi^2$/dof (XIS). (3) p-value (XIS). (4) $\chi^2$/dof (HXD). (5) p-value (HXD).

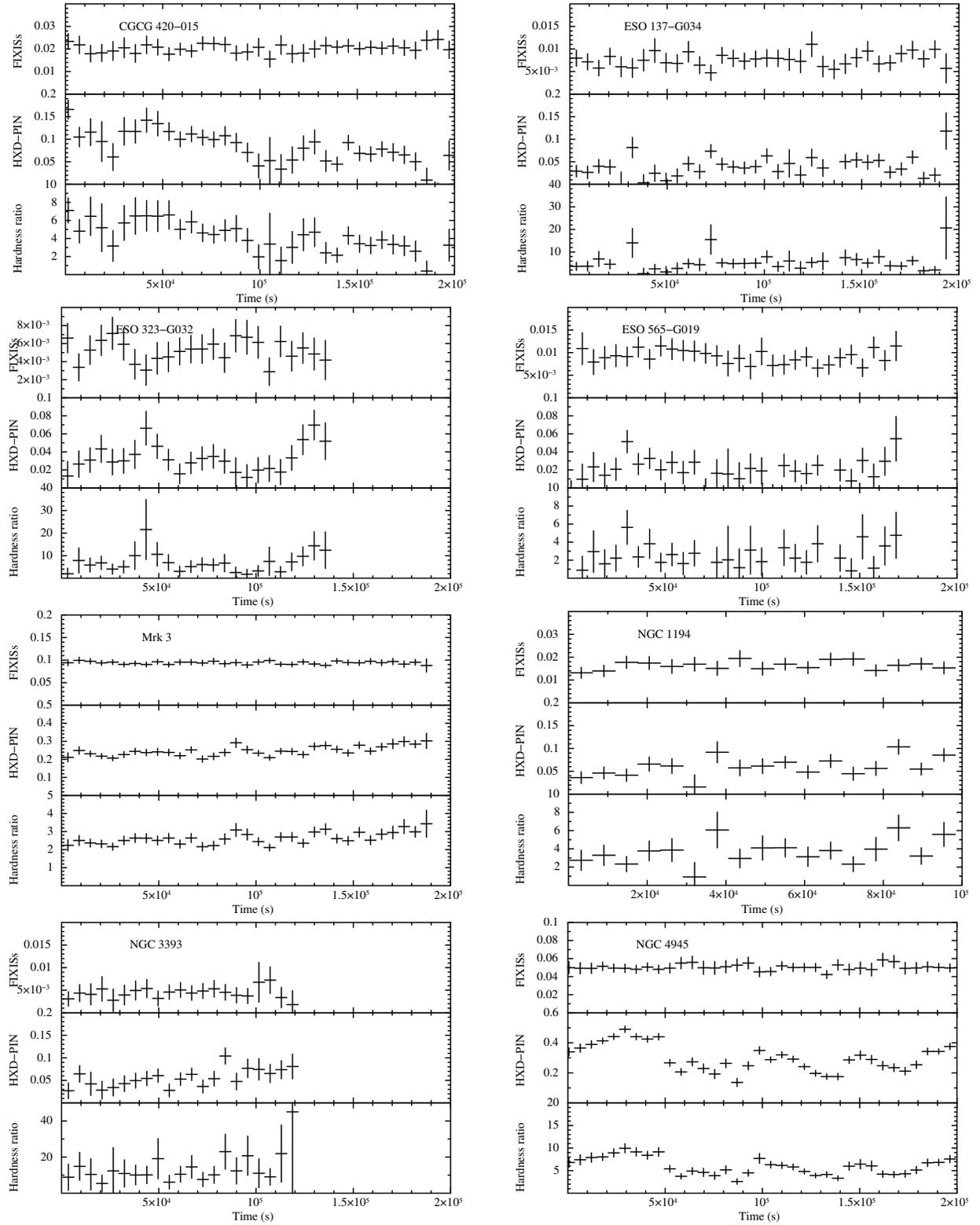

FIG. 1.— Background-subtracted light curves in units of counts $s^{-1}$. Upper panel: *Suzaku*/FIXISs (2–10 keV). Middle panel: *Suzaku*/HXD-PIN (16–40 keV). Bottom panel: the hardness ratio between them.

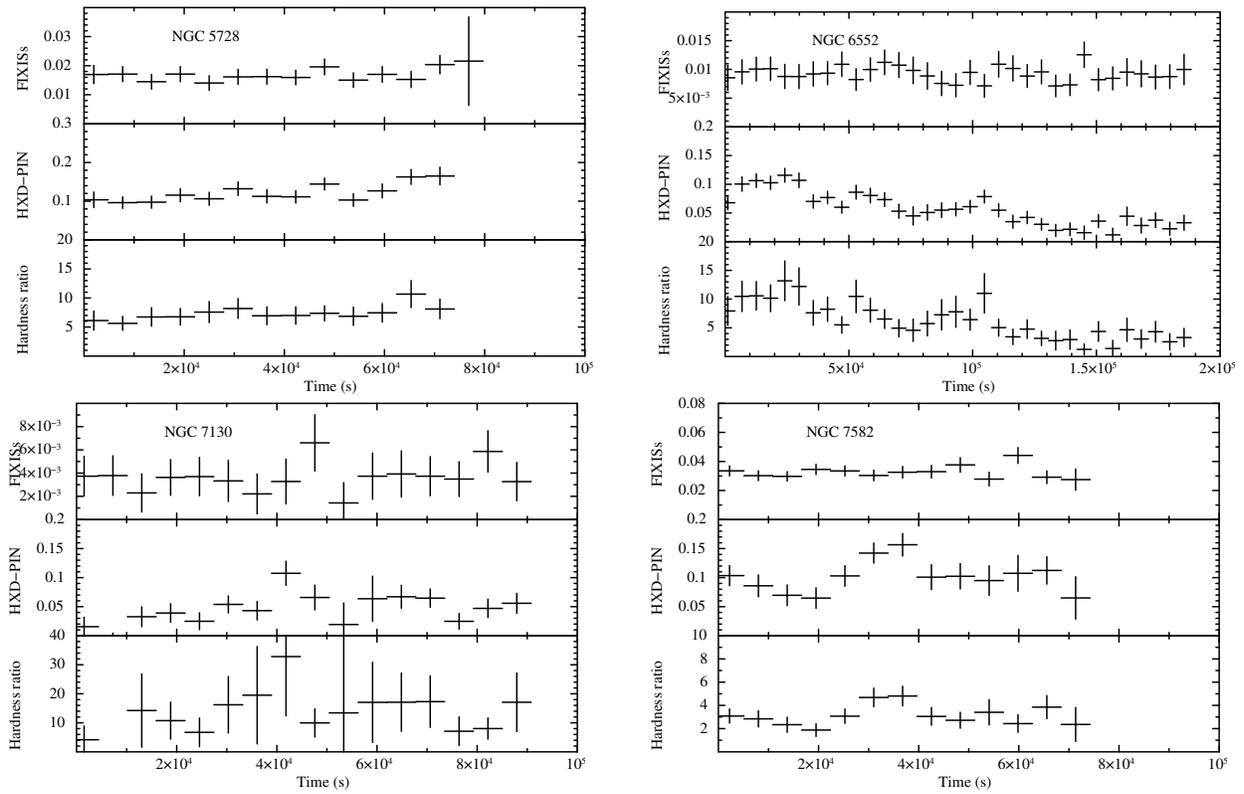

Fig. 1.— (Continued)

## 3. LIGHT CURVES AND SPECTRA

### 3.1. *Light curves*

Figure 1 shows the background-subtracted light curves of the FIXISs (2–10 keV) and the HXD-PIN (16–40 keV), together with their hardness ratio (16–40 keV/2–10 keV), for all targets. To minimize any systematic uncertainties caused by the orbital motion of the satellite, we set the time bin size to the orbital period (5760 sec). To check possible time variability, we perform a $\chi^2$ test to each light curve assuming a null hypothesis of a constant flux (Table 3). We find that the FIXIS light curves are consistent with being constant for all targets. On the other hand, the HXD-PIN light curves of CGCG 420–015, NGC 4945, and NGC 6552 show significant time variability at > 99% confidence levels. Similar short-term variability of NGC 4945 was also reported by Yaqoob (2012) and Puccetti et al. (2014). We have confirmed that the shape of the HXD-PIN spectra of these 3 targets did not vary over statistical errors depending on flux. Thus, in this paper we analyze the spectra averaged over the whole observation for all targets.

### 3.2. *Spectra*

We perform a simultaneous fit to the *Suzaku*/BIXIS (0.5–8.0 keV), *Suzaku*/FIXISs (2–10 keV), *Suzaku*/HXD (16–40 keV: widest case) and 70-month averaged *Swift*/BAT (14–100 keV) spectra. Because our targets are very faint at soft e energy bands, we only utilize the BIXIS below 2 keV, which has much superior sensitivity with respect to the FIXISs. The 1.7–1.9 keV of the BIXIS spectra is excluded to avoid systematic uncertainties in the energy response around the Si K-edge. We conservatively decide not to use the *Swift*/BAT spectra above 100 keV, where the signal to noise ratios are low for most of the targets. In the following, we uniformly apply analytic models (Section 3.2.1) and numerical torus models by Ikeda et al. (2009) (Section 3.2.2) to these spectra.

#### 3.2.1. *Baseline Models*

First, we apply two analytic models (Baseline1 and Baseline2) often adopted to represent the broadband spectra of obscured AGNs (e.g., Schurch et al. 2002; Matt et al. 2004; Bianchi et al. 2005; Piconcelli et al. 2007; Ueda et al. 2007; Comastri et al. 2010; Kawamuro et al. 2016a; Tanimoto et al. 2016). The models are basically composed of an absorbed direct component, an unabsorbed scattered component, a reflection component and narrow Fe K emission lines. Other emission lines and/or optically-thin thermal components are added when required from the data. The difference between the Baseline1 and Baseline2 models are that Compton scattering for the direct components is ignored in the former but is taken into account in the latter. Note that it is not trivial which of the two models is more physically reasonable, depending on the geometry considered (Section 5.2).

In the XSPEC terminology, the Baseline1 and Baseline2 models are expressed as

$$
\begin{aligned}
\text{Baseline1} = {} & \text{const1} * \text{phabs} \\
& * (\text{const2} * \text{zphabs} * \text{zpowerlw} * \text{zhighect} \\
& + \text{const3} * \text{zpowerlw} * \text{zhighect} \\
& + \text{zphabs} * \text{pexrav} + \text{zgausses} + \text{apec(s)}) \quad (1)
\end{aligned}
$$

$$
\begin{aligned}
\text{Baseline2} = {} & \text{const1} * \text{phabs} \\
& * (\text{const2} * \text{zphabs} * \text{cabs} * \text{zpowerlw} * \text{zhighect} \\
& + \text{const3} * \text{zpowerlw} * \text{zhighect} \\
& + \text{zphabs} * \text{pexrav} + \text{zgausses} + \text{apec(s)}) \quad (2)
\end{aligned}
$$

0. We multiply a cross normalization factor (const1) to reflect differences in the absolute flux calibration among the instruments. We adopt the *Swift*/BAT data as reference, whose const1 value is set to unity. We also fix that of *Suzaku*/FIXISs at unity, by assuming that there is no relative calibration error between *Suzaku*/FIXISs and *Swift*/BAT[2]. The cross normalization of *Suzaku*/BIXIS is left as a free parameter ($N_{\text{BIXIS}}$), whereas that of the *Suzaku*/HXD is set to 1.16 (1.18) for XIS (HXD) nominal position observations according to the calibration results obtained using the Crab Nebula[3]. We consider Galactic absorption (phabs) whose column density is fixed at a value estimated from HI map for each target (Kalberla et al. 2005).

1. The first term represents the absorbed direct component. The intrinsic continuum is modeled by a power law (zpowerlw) with an exponential cutoff (zhighect). In the Baseline1, we only consider photoelectric absorption (zphabs) as is the case in Kawamuro et al. (2016a) for Compton thin AGNs. In the Baseline2 model, we also consider Compton scattering out of the line of sight (cabs) whose hydrogen column density ($N_{\text{H}}^{\text{Dir}}$) is linked to that of zphabs. Since we cannot constrain the cutoff energy[4]. with our data, we fix it at 360 keV, the value adopted in the Ikeda torus model (Ikeda et al. 2009). We have confirmed that the choice of this value within a typical range observed in local AGNs (100–500 keV) does not affect significantly our spectral parameters. The factor $N_{\text{Suzaku}}$ (const2) is multiplied to take into account possible time variation of the direct-component flux during the *Suzaku* observation with respect to 70-month averaged *Swift*/BAT flux. Note that this constant factor is not multiplied to the scattered and reflection components (the second and third terms), assuming that they did not vary between the *Suzaku* and *Swift*/BAT observations as the size of the reflector has most likely a pc scale (Kawamuro et al. 2016a). To avoid unrealistic results, we limit the $N_{\text{Suzaku}}$ value within a range of 0.2–5.0, according to the results obtained by Kawamuro et al. (2016a) for Compton-thin AGNs.

---

[2] In reality, a possible error may exist, which is coupled to the $N_{\text{Suzaku}}$ parameter introduced below.
[3] http://www.astro.isas.jaxa.jp/suzaku/doc/suzakumemo/suzakumemo-2008-06.eps
[4] This corresponds to an e-folding energy ($E_{\text{f}}$) in the zhighect model when $E_{\text{c}}$ is set to be zero

2. The second term represents the unabsorbed scattered component. We multiply the scattering fraction $f_{\rm scat}$ (const3) to a cutoff power-law with the same photon index and normalization as the first term.

3. The third term represents the reflection component. We adopt the analytic code (pexrav) by Magdziarz & Zdziarski (1995), which calculates a reflected continuum from cold, optically-thick matter. The reflection strength relative to the direct component is defined by $R = \Omega/2\pi$ ($\Omega$ is the solid angle of the reflector)[5]. To avoid unphysical parameters, we impose an upper limit of $R = 2$, corresponding to the extreme case where the nucleus is covered by the reflector in all directions. The inclination angle to the reflector is fixed at 60 degrees. The photon index and normalization of the incident cutoff power law are linked to those of the first term. We consider photoelectric absorption to this component with a hydrogen column density of $N_{\rm H}^{\rm Ref}$, independently of $N_{\rm H}^{\rm Dir}$, if it is found to be significantly required (i.e., $N_{\rm H}^{\rm Ref} > 0$) over a 99% confidence limit.

4. The fourth term represents emission lines of Fe K$\alpha$ around 6.40 keV and K$\beta$ around 7.06 keV (zgauss). We fix these line width at 10 eV for both lines. We set the central energy $E_{\rm FeK\alpha}$ as a free parameter while we fix the central energy $E_{\rm FeK\beta}$ at 7.06 keV. In the case of Mrk 3 and NGC 4945, we also include other emission lines that are significantly detected in previous *Suzaku* papers (Sections 4.4 and 4.7).

5. The fifth term represents emission from optically-thin thermal plasmas (apec) in the host galaxy (Smith et al. 2001). We test whether inclusion of an apec component to the above model improves the fit. We adopt it if the improvement is found to be significant at a 99% confidence limit (i.e., $\Delta\chi^2 \leq -9.21$, which correspond to the 99% limit of the $\chi^2$ distribution with a degree of freedom of 2). At maximum two apec components with different temperatures are found to be required from the data.

These analytic models reproduce the *Suzaku* and *Swift*/BAT spectra of all targets reasonably well ($\chi^2$/dof $\leq 1.3$). Tables 4 and 5 summarizes the best-fit parameters of the Baseline1 and Baseline2 models, respectively. In general, we obtain slightly larger best-fit photon indices with the Baseline2 model than with the Baseline1 model because of the Klein-Nishina decline of the Compton scattering cross section. The Baseline2 model gives a larger best-fit intrinsic luminosity (Section 5.2) and accordingly a smaller scattering fraction than the Baseline1 model. The comparison of intrinsic luminosity between the two models is discussed in Section 5.2. In most cases, the spectral parameters obtained with the two models are consistent, except for the brightest sources (Mrk 3 and NGC 4945). Figures 2 and 3 (left column) plot the unfolded spectra and best-fit Baseline2 model, respectively, in units of $EF_E$ ($F_E$ is the energy flux density at an energy of $E$). Figure 4 plots the folded spectra in the energy around Fe K$\alpha$ lines. Table 8 lists the observed fluxes in the 0.5–2.0 keV, 2–10 keV and 10–50 keV, the intrinsic luminosities (i.e., de-absorbed luminosities not including reflection component) in the 0.5–2.0 keV, 2–10 keV and 10–50 keV, the Fe K$\alpha$ luminosity and the Eddington ratio, based on the best-fit Baseline2 model. We define the Eddington luminosity as $L_{\rm Edd} = 1.26 \times 10^{38} M_{\rm BH}/M_\odot$ erg s$^{-1}$ and adopt a bolometric correction factor of 20 (Vasudevan & Fabian 2009) from the 2–10 keV luminosity averaged over 70 months (i.e, based on the *Swift*/BAT normalization).

### 3.2.2. *Torus Models*

Next, we apply Monte-Carlo based numerical spectral models from uniform-density tori developed by Ikeda et al. (2009) (we call Ikeda1 and Ikeda2). Since this model is available only above 1 keV, we limit the energy range to 1–100 keV in the spectral fit. The Ikeda torus model assumes a nearly spherical geometry with two conical holes along the polar axis, and has three parameters that can be set free: the hydrogen column density along the equatorial plane $N_{\rm H}^{\rm Equ}$ (within a range of $10^{22}$–$10^{25}$ cm$^{-2}$), the inclination angle $\theta_{\rm incl}$ (1–89 degrees) and the half-opening angle of the torus $\theta_{\rm open}$ (10–70 degrees). The ratio between the inner radius $R_{\rm in}$ and the outer radius $R_{\rm out}$ is fixed at 0.01. As the incident spectrum, a power law with an exponential cutoff of 360 keV is assumed, whose photon index is a free parameter within a range of 1.5–2.5.

In the XSPEC terminology, these models[6] are represented as follows:

$$
\begin{aligned}
\mathsf{Ikeda1} = &\; \mathsf{const1} * \mathsf{phabs} \\
& * (\, \mathsf{const2} * \mathsf{torusabs} * \mathsf{zpowerlw} * \mathsf{zhighect} \\
& + \mathsf{const3} * \mathsf{zpowerlw} * \mathsf{zhighect} \\
& + \mathsf{mtable\{refl\_all\_torus.fits\}} * \mathsf{zpowerlw} * \mathsf{zhighect} \\
& + \mathsf{const4} * \mathsf{atable\{refl\_fe\_torus.fits\}} \\
& + \mathsf{zgausses} + \mathsf{apec(s)})
\end{aligned} \quad (3)
$$

$$
\begin{aligned}
\mathsf{Ikeda2} = &\; \mathsf{const1} * \mathsf{phabs} \\
& * (\, \mathsf{const2} * \mathsf{zphabs} * \mathsf{cabs} * \mathsf{zpowerlw} * \mathsf{zhighect} \\
& + \mathsf{const3} * \mathsf{zpowerlw} * \mathsf{zhighect} \\
& + \mathsf{mtable\{refl\_all\_torus.fits\}} * \mathsf{zpowerlw} * \mathsf{zhighect} \\
& + \mathsf{const4} * \mathsf{atable\{refl\_fe\_torus.fits\}} \\
& + \mathsf{zgausses} + \mathsf{apec(s)})
\end{aligned} \quad (4)
$$

These models are composed of five components similar to the case of the baseline models. (0) Same as the baseline models. (1) The absorbed direct component. (2) The scattered component. (3) The reflection component from the torus. (4) The Fe K$\alpha$ emission line from the torus, whose relative normalization $N_{\rm FeK\alpha}$ (const4) to the reflection continuum is set free. The reason why we introduce this factor is to take into account the uncertainties caused by the assumption of geometry and a possible non-Solar abundance of Fe. Other emission lines than Fe Kα are separately modeled by gaussians. (5) The optically-thin thermal component(s) of the host galaxy if required. The temperature and normalization are fixed at the values obtained with the Baseline2 model, which are difficult to constrain without using the low energy below 1 keV. The photon index and power-law normalization are linked together among the direct, scattered, reflection and Fe Kα line components.

---

[5] We set $R < 0$ to reproduce only the reflection component without the direct component

[6] Recently a "mtable" model to reproduce the reflected continuum has been released, which replaces old "atable" models used in previous works

On the basis of this model, we consider two cases (Ikeda1 and Ikeda2) with different settings for spectral parameters. In the Ikeda1 model, we assume an ideal case where the line-of-sight column density ($N_\mathrm{H}^\mathrm{LOS}$) is purely determined by the torus geometrical parameters, the hydrogen column density along the equatorial plane, the inclination angle, and the half-opening angle, which are set to be free. In the case of $\theta_\mathrm{incl} \geq \theta_\mathrm{open}$, it is determined as

$$N_\mathrm{H}^\mathrm{LOS} = \frac{r(\cos\theta_\mathrm{incl} - \cos\theta_\mathrm{open}) + \sin(\theta_\mathrm{incl} - \theta_\mathrm{open})}{(1-r)(r\cos\theta_\mathrm{incl} + \sin(\theta_\mathrm{incl} - \theta_\mathrm{open}))} N_\mathrm{H}^\mathrm{Equ}. \quad (5)$$

In the Ikeda2 model, the line-of-sight hydrogen column density is an independent free parameter decoupled from the torus parameters. This is based on an idea of clumpy tori, where the hydrogen column density along a single direction can vary from the averaged hydrogen column density of the torus. Accordingly, we replace torusabs with cabs*phabs. Instead, we fix the inclination angle at 70 degrees (the upper boundary of the half-opening angle) except for Mrk 3 (Section 4.5) and set the lower limit of $\log N_\mathrm{H}^\mathrm{Equ}/\mathrm{cm}^{-2} = 23.5$, since we find it difficult to constrain the three geometrical parameters in this case.

Tables 6 and 7 summarize the best-fit parameters of the Ikeda1 and Ikeda2 models, respectively. We note that the fit with the Ikeda1 model for Mrk 3 and NGC 6552 is not good, which is improved with the Ikeda2 model. Figures 2 and 3 (right columns) plot the unfolded spectra and corresponding best-fit models, respectively, based on the Ikeda2 model. Table 9 lists the observed fluxes and intrinsic luminosities in the 0.5–2.0 keV, 2–10 keV and 10–50 keV, the Fe Kα luminosities, and the Eddington ratios (calculated in the same way as described in Section 3.2.1), based on the best-fit Ikeda2 model.

## 4. RESULTS OF INDIVIDUAL SPECTRA

We summarize the main results of the spectral fits for each target. In most cases, we obtain consistent results among the different models 1applied (i.e., Baseline1, Baseline2, Ikeda1 and Ikeda2) in terms of the line-of-sight hydrogen column density and photon index. Thus, we mainly refer to the values obtained with the Ikeda2, which we believe gives the most physically realistic picture (Section 5.2). We also compare our results with previous studies[7] of *Chandra*, *XMM-Newton*, *Suzaku* and/or *NuSTAR* data using the MYTorus model (Murphy &

[7] To avoid any confusions, in this paper we do not refer to previous results using the Torus model by Brightman & Nandra (2011a,b), where there was an inconsistency between the assumed geometry and output spectra (Liu & Li 2015), Nevertheless, the Torus model works quite well in reproducing actual CTAGN spectra (Buchner et al., in preparation). The column densities obtained by Ricci et al. (2015) with the Torus model are consistent with our Ikeda2 model results except for NGC 4945.

Yaqoob 2009). The MYTorus model assumes a uniform-density torus of a donut-like shape with an opening angle fixed at 60 degrees, and have two geometrical free parameters: the inclination and equatorial column density. For convenience, we convert the equatorial column density into the line-of-sight column density ($N_\mathrm{H}^\mathrm{Dir}$) whenever we refer to results with the MYTorus model.

### 4.1. *CGCG 420–015*

All the models with two apec components well reproduce the broadband spectra. We detect Fe Kα ($EW = 0.26^{+0.05}_{-0.10}$ keV) and Kβ ($EW = 0.05^{+0.12}_{-0.01}$ keV) emission lines. We obtain $N_\mathrm{H}^\mathrm{Dir} = 1.23^{+1.65}_{-0.27} \times 10^{24}$ cm$^{-2}$, $\Gamma = 2.34(>2.08)$ with the Ikeda2 model. Severgnini et al. (2011) obtained $N_\mathrm{H}^\mathrm{Dir} = 1.46^{+0.07}_{-0.11} \times 10^{24}$ cm$^{-2}$ and $\Gamma = 2.40^{+0.5}_{-0.5}$ analyzing the *Suzaku* data combined with *Swift*/BAT 54-month spectra. Our results are fully consistent with their results.

### 4.2. *ESO 137–G034*

All the models with one apec component well reproduce the broadband spectra. We detect Fe Kα ($EW = 1.03^{+1.48}_{-0.66}$ keV) and Kβ ($EW = 0.21^{+0.11}_{-0.13}$ keV) emission lines. We obtain $N_\mathrm{H}^\mathrm{Dir} = 1.30^{+0.37}_{-0.34} \times 10^{24}$ cm$^{-2}$ and $\Gamma = 2.11^{+0.34}_{-0.32}$ with the Ikeda2 model. Comastri et al. (2010) reported $N_\mathrm{H}^\mathrm{Dir} = 1.0 \times 10^{25}$ cm$^{-2}$ (fixed) and $\Gamma = 1.58^{+0.16}_{-0.20}$ using the *Suzaku* data with applying a reflection-dominated model that is different from our models. As described in Section 5.5, the X-ray luminosity we obtain is consistent with that expected from the mid-infrared luminosity.

### 4.3. *ESO 323–G032*

All the models with one apec component well reproduce the broadband spectra. A Fe Kα line ($EW = 1.27^{+0.62}_{-0.48}$ keV) is detected. We obtain $N_\mathrm{H}^\mathrm{Dir} = 2.24^{+5.29}_{-0.74} \times 10^{24}$ cm$^{-2}$ and $\Gamma = 2.30(>2.02)$ with the Ikeda2 model. Comastri et al. (2010) obtained $N_\mathrm{H}^\mathrm{Dir} = 1.0 \times 10^{25}$ cm$^{-2}$ (fixed) and $\Gamma = 1.85^{+0.47}_{-0.40}$ using the *Suzaku* data with the reflection-dominant model. Again, our X-ray luminosity is consistent with the mid-infrared luminosity (Section 5.5).

### 4.4. *ESO 565–G019*

All the models with one apec component well reproduce our broadband spectra. We detect Fe Kα ($EW = 1.16^{+0.32}_{-0.39}$ keV) and Kβ ($EW = 0.15^{+0.05}_{-0.11}$ keV) lines. We obtain $N_\mathrm{H}^\mathrm{Dir} = 2.17^{+0.50}_{-0.30} \times 10^{24}$ cm$^2$ and $\Gamma = 1.95^{+0.50}_{-0.23}$ with the Ikeda2 model. Gandhi et al. (2013) obtained a line-of-sight column density of $N_\mathrm{H}^\mathrm{Dir} = 4.4(>2.9) \times 10^{24}$ cm$^{-2}$ and $\Gamma = 1.72^{+0.41}_{-0.27}$ using the *Suzaku* data with the MYTorus model. Whereas our $\Gamma$ value is consistent with the result of Gandhi et al. (2013), the $N_\mathrm{H}^\mathrm{Dir}$ values are subtly different. This is most probably due to the difference in the assumed geometry of the two torus models.

### 4.5. *Mrk 3*

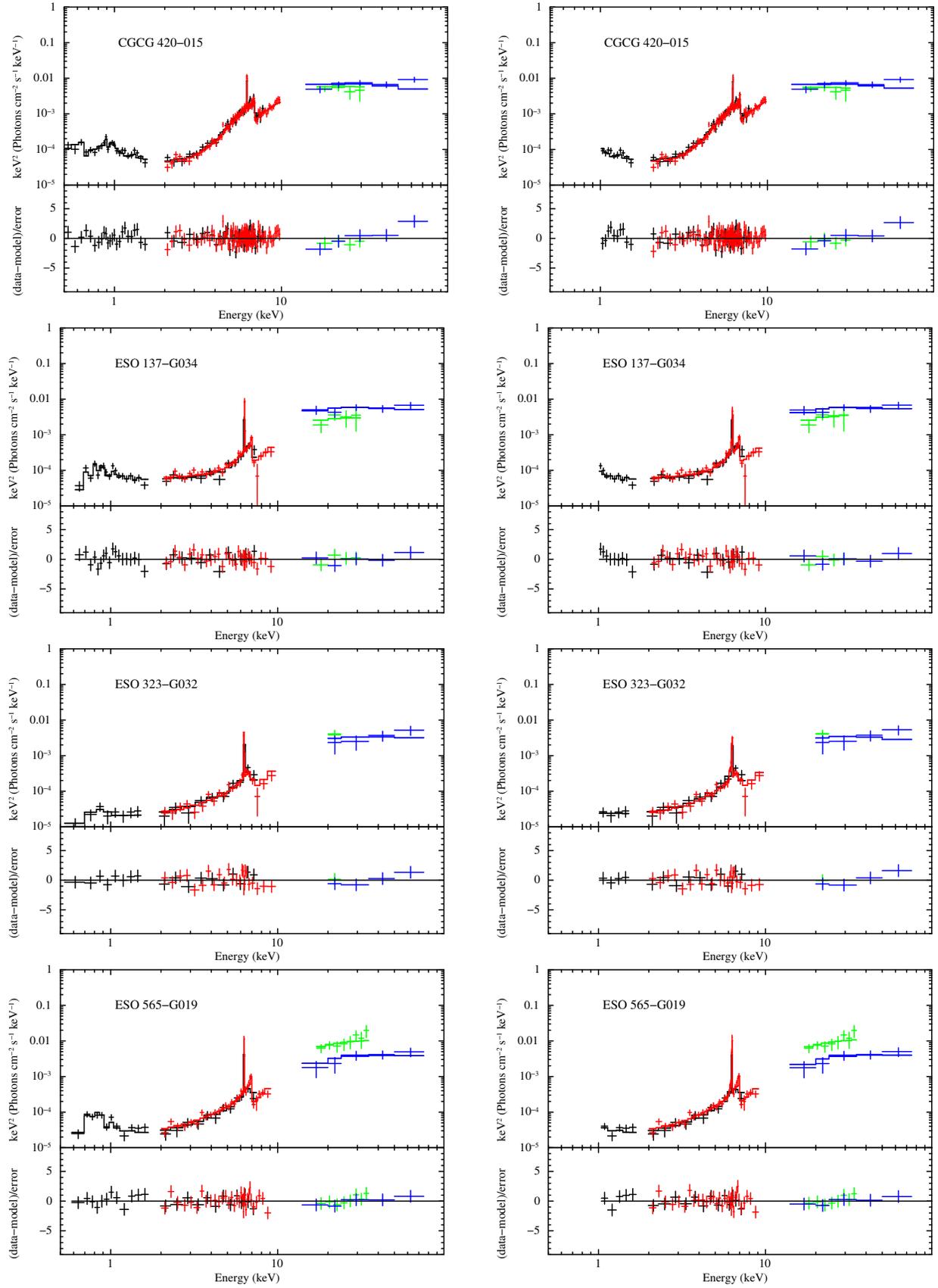

FIG. 2.— The unfolded spectra fitted with the Baseline2 model (left) and the Ikeda2 model (right). In the upper panels, the unfolded spectra of *Suzaku*/BIXIS (black crosses), *Suzaku*/FIXISs (red crosses), *Suzaku*/HXD-PIN (green crosses) and *Swift*/BAT (blue crosses) are represented. The solid curves show the best fit model. In the lower panels, the residuals are shown.

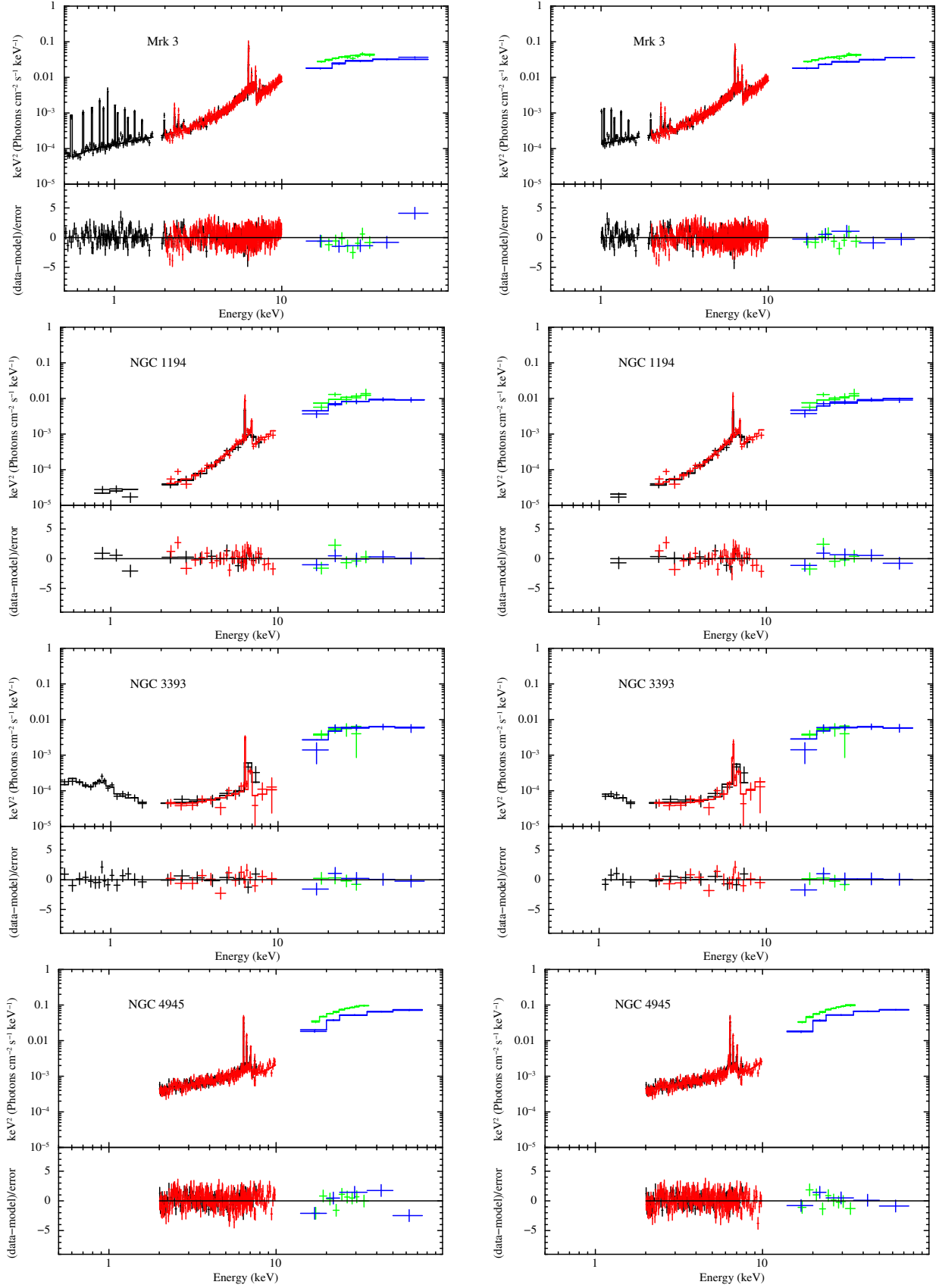

Fig. 2.— (Continued)

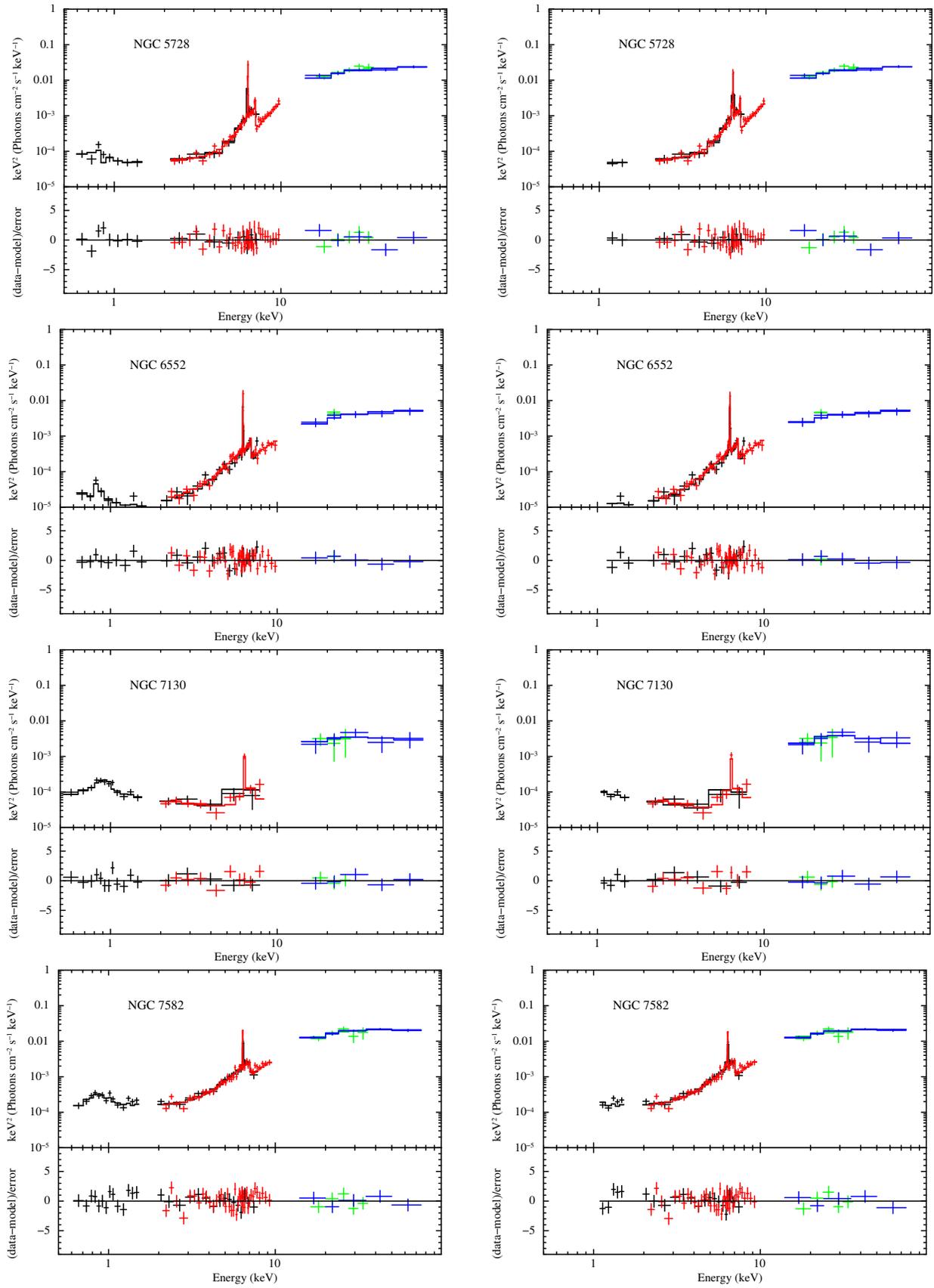

Fig. 2.— (Continued)

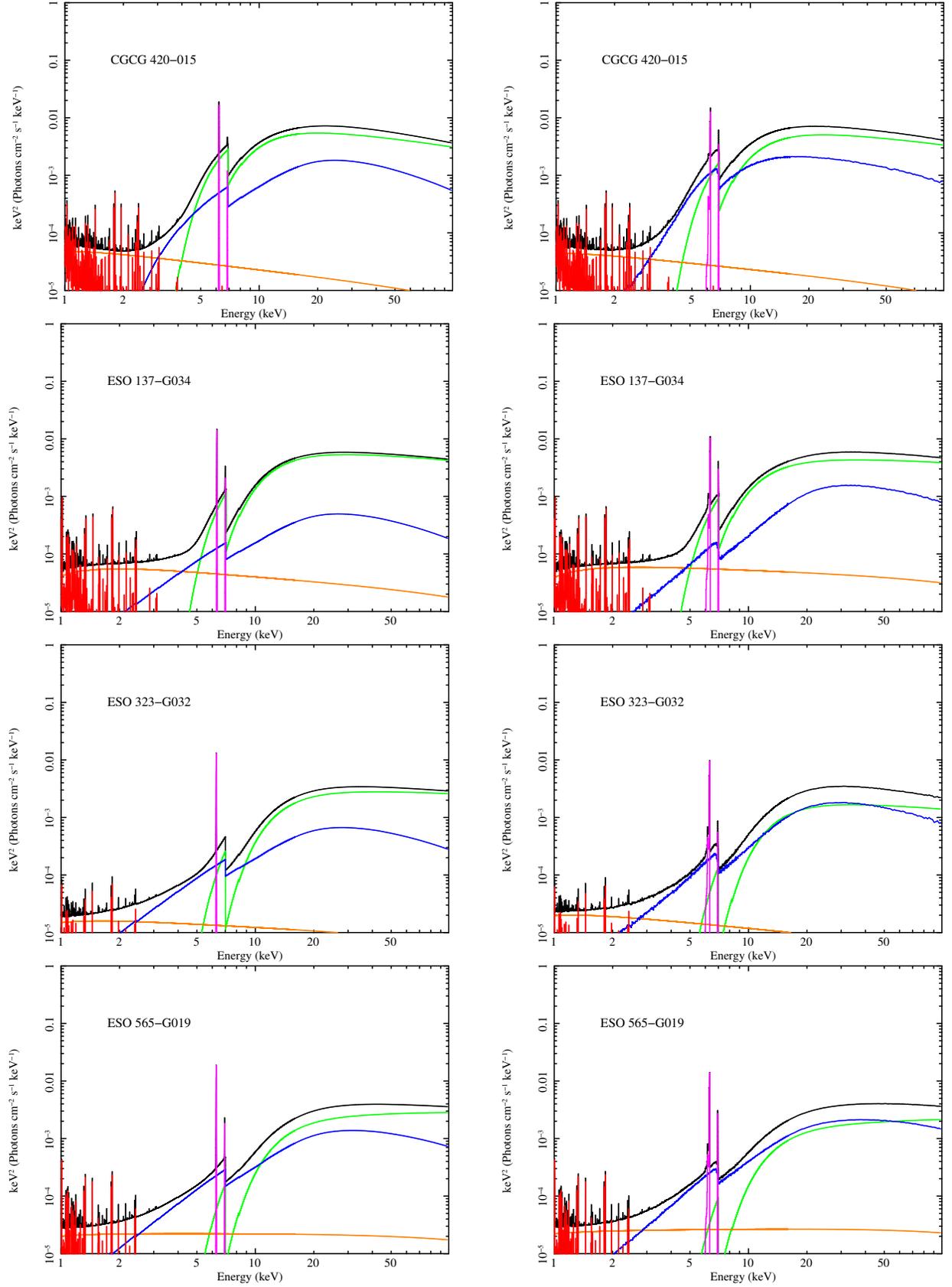

Fig. 3.— The best fitting spectral models with the Baseline2 (left) and the Ikeda2 (right). The black, green, orange, blue, magenta and red lines represent the total, direct component, scattered component, reflection component, emission lines and emission from an optically thin thermal plasma, respectively.

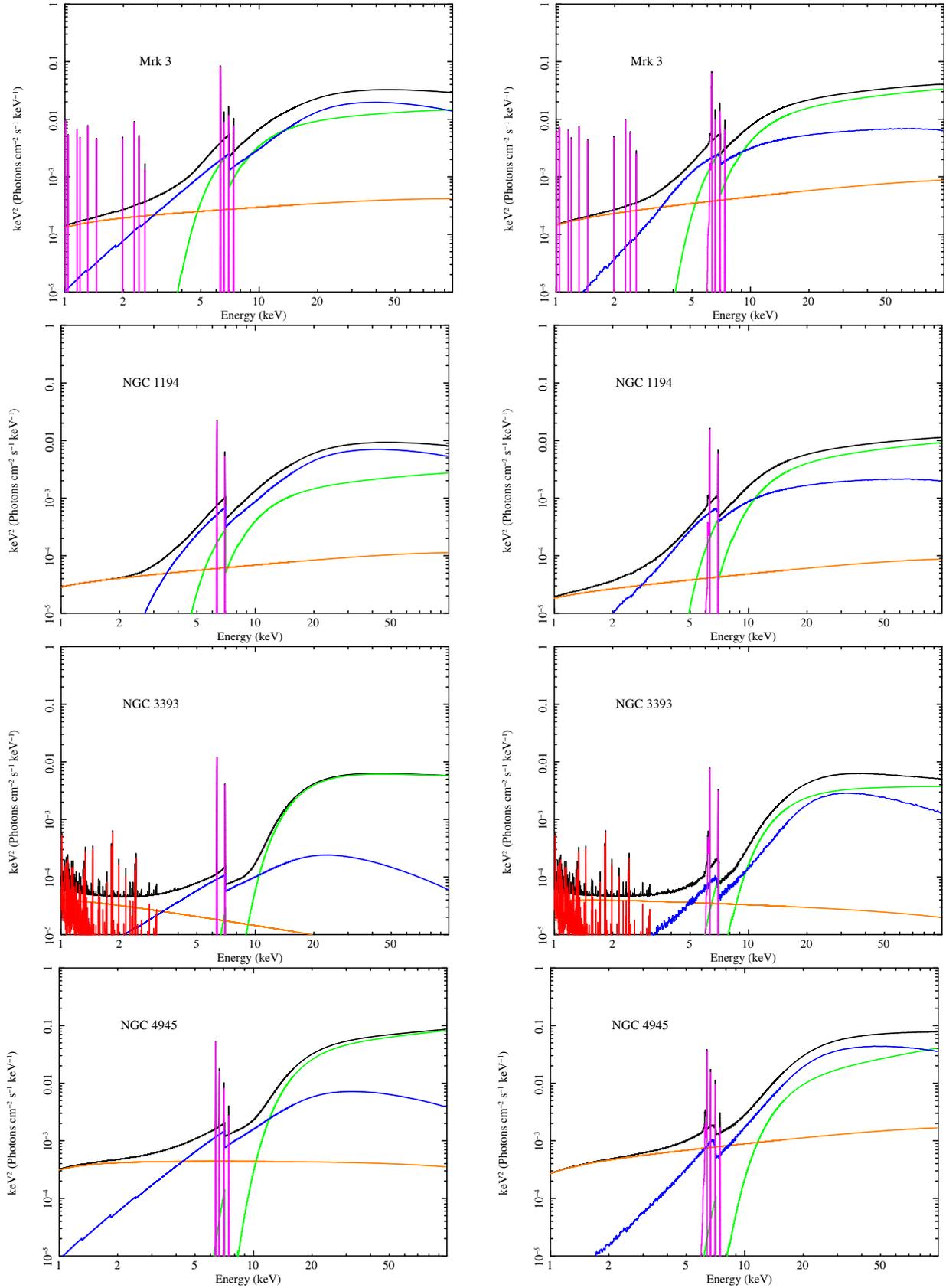

Fig. 3.— (Continued)

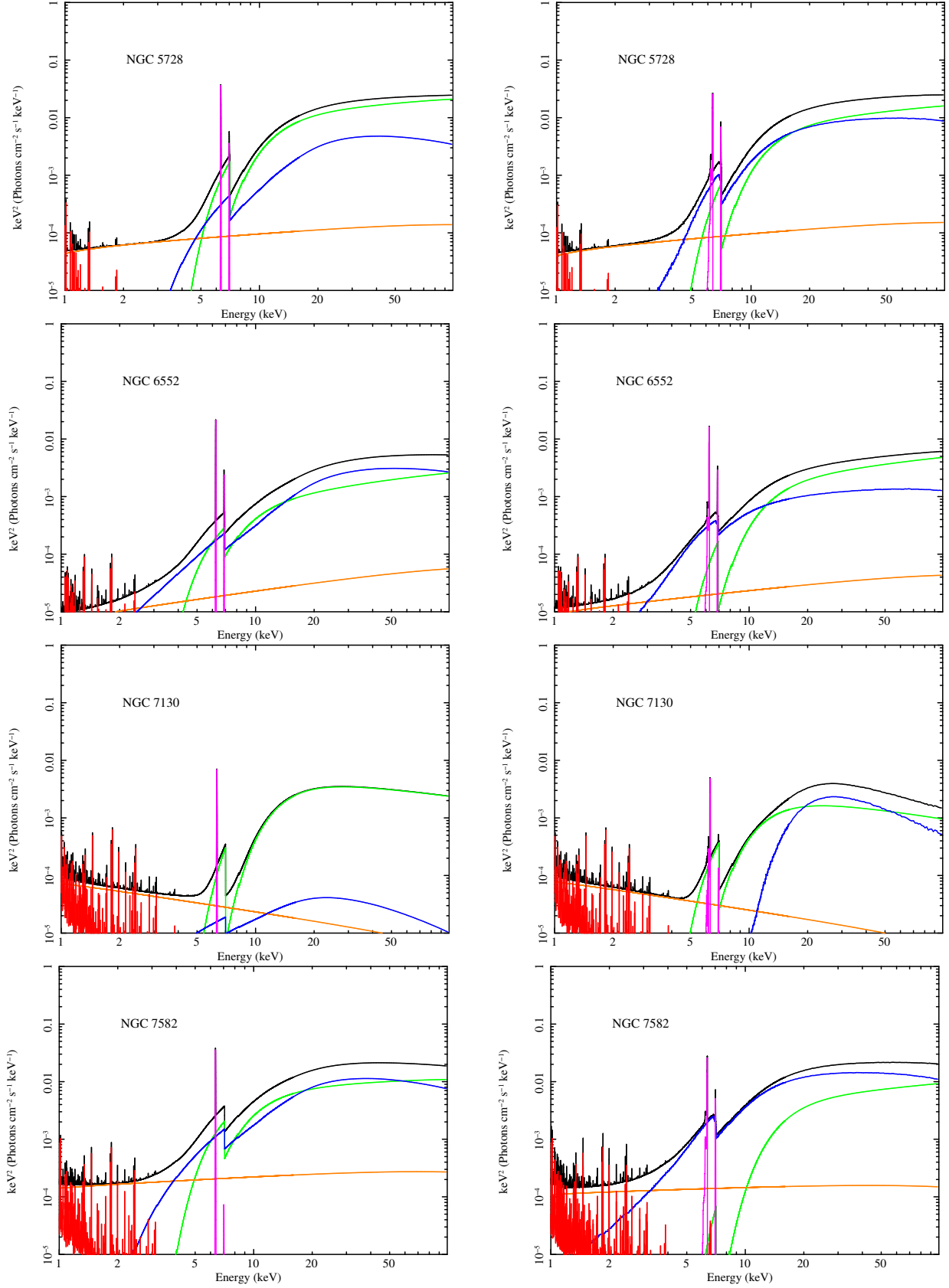

Fig. 3.— (Continued)

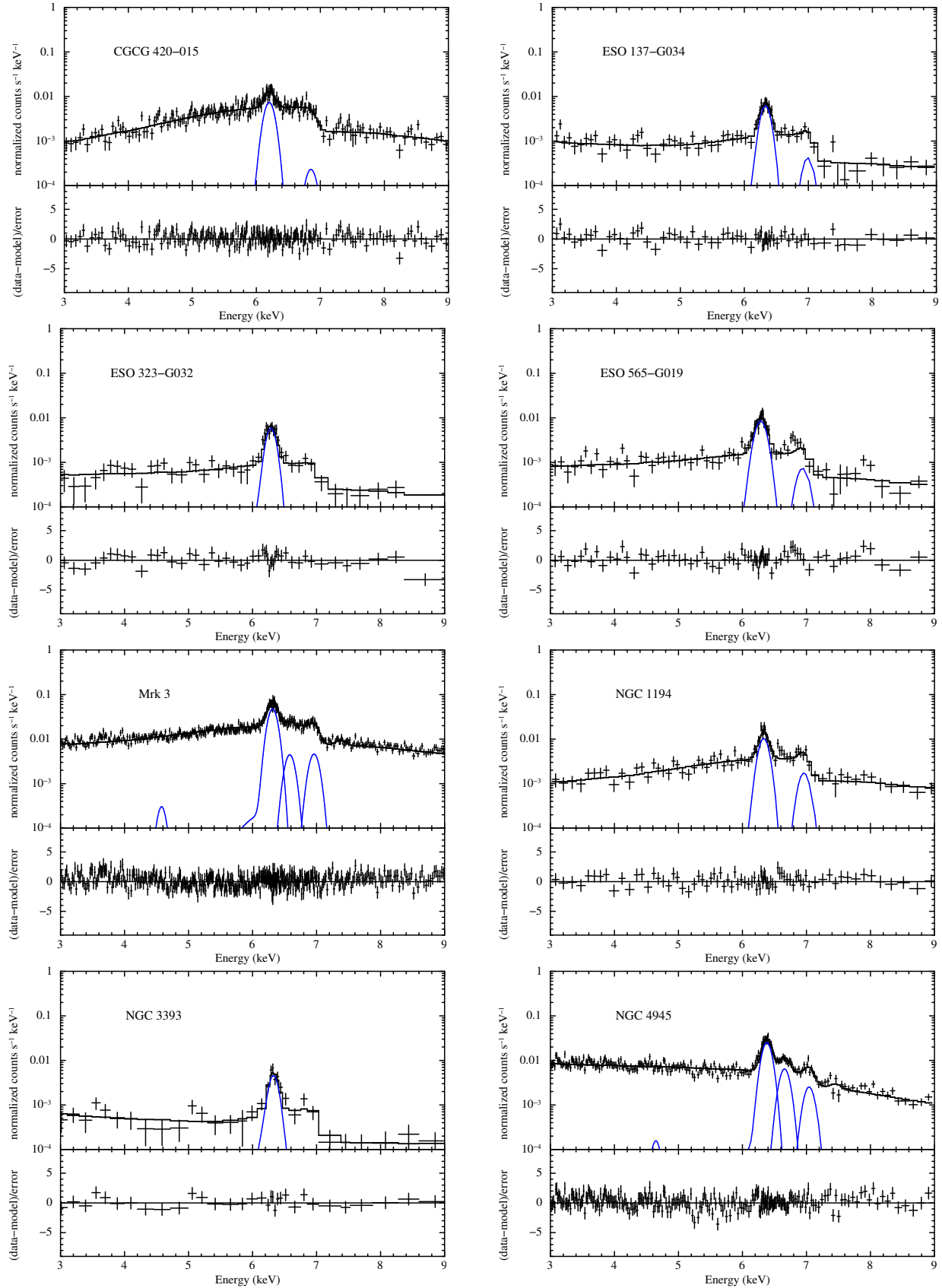

FIG. 4.— The observed spectra of *Suzaku*/FIXISs in the 3–9 keV band folded with the energy response. The black crosses and the solid curve represent the data and the best-fit model (Baseline2), respectively. The blue curves show the contribution of the Fe K emission lines; the weak line feature at 4–5 keV is a "Si-K escape" peak of the Fe K$\alpha$ line caused by the instrumental response of the FIXISs.

<pre>                                                                                          17</pre>

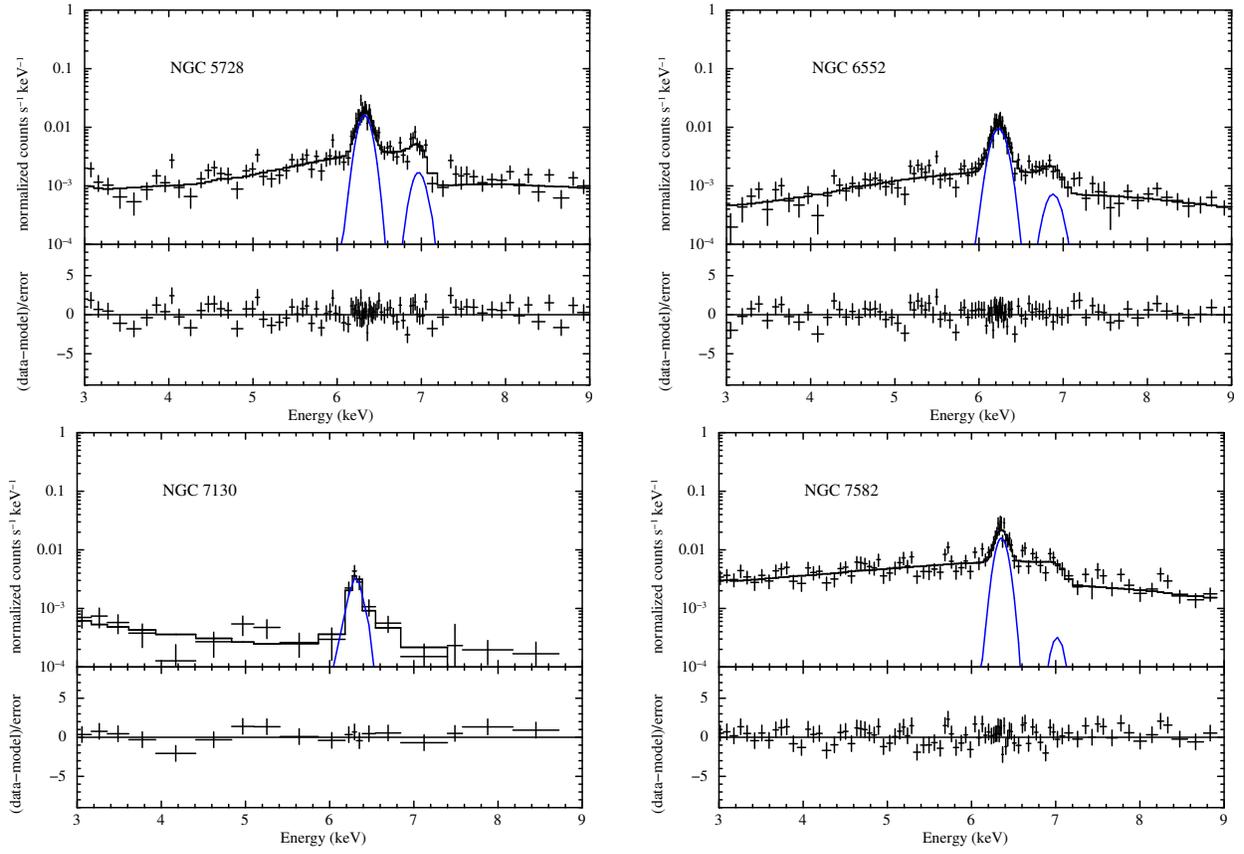

Fig. 4.— (Continued)

TABLE 4
Best fit parameters with the Baseline1 model

| (1) Galaxy Name | (2) $N_{\rm BIXIS}$ (10) $k_B T_1$ | (3) $N_{\rm Suzaku}$ (11) $N_{\rm Apec1}$ | (4) $N_{\rm H}^{\rm Dir}$ (12) $k_B T_2$ | (5) $\Gamma$ (13) $N_{\rm Apec2}$ | (6) $N_{\rm Dir}$ (14) $E_{\rm FeK\alpha}$ | (7) $f_{\rm scat}$ (15) $I_{\rm FeK\alpha}$ | (8) $N_{\rm H}^{\rm Ref}$ (16) $E_{\rm FeK\beta}$ | (9) $R_{\rm Ref}$ (17) $I_{\rm FeK\beta}$ | $\chi^2$/dof |
|---|---|---|---|---|---|---|---|---|---|
| CGCG 420−015 | $1.04^{+0.05}_{-0.05}$ $0.18^{+0.05}_{-0.04}$ | $0.59^{+0.19}_{-0.13}$ $0.79^{+0.68}_{-0.37}$ | $0.91^{+0.14}_{-0.12}$ $0.99^{+0.10}_{-0.10}$ | $2.36^{+0.05}_{-0.09}(>2.19)$ $0.37^{+0.10}_{-0.09}$ | $2.03^{+1.58}_{-0.95}$ $6.39^{+0.01}_{-0.01}$ | $0.29^{+0.22}_{-0.12}$ $1.08^{+0.15}_{-0.15}$ | $0.09^{+0.05}_{-0.04}$ $7.06$ (fixed) | $0.54^{+0.37}_{-0.20}$ $0.07^{+0.15}_{-0.19}(<0.19)$ | 155.7/154.0 |
| ESO 137−G034 | $0.91^{+0.09}_{-0.14}$ $0.66^{+0.14}_{-0.09}$ | $0.37^{+0.20}_{-0.14}$ $0.61^{+0.15}_{-0.14}$ | $1.43^{+0.42}_{-0.37}$ ... | $2.15(>1.78)$ ... | $1.06^{+2.79}_{-0.79}$ $6.39^{+0.01}_{-0.01}$ | $0.67^{+1.50}_{-0.50}$ $0.88^{+0.15}_{-0.12}$ | ... $7.06$ (fixed) | $0.14^{+0.10}_{-0.08}$ $0.11^{+0.09}_{-0.09}$ | 53.5/64.0 |
| ESO 323−G032 | $1.05^{+0.14}_{-0.13}$ $0.74^{+0.63}_{-0.44}$ | $1.07^{+1.68}_{-0.63}$ $0.06^{+0.05}_{-0.05}$ | $1.98^{+0.58}_{-0.72}$ ... | $2.08(>1.54)$ ... | $0.43^{+2.15}_{-0.40}$ $6.39^{+0.02}_{-0.02}$ | $0.44^{+4.89}_{-0.39}$ $0.82^{+0.12}_{-0.12}$ | ... $7.06$ (fixed) | $0.39(>0.13)$ $0.00(<0.01)$ | 42.1/37.0 |
| ESO 565−G019 | $0.89^{+0.09}_{-0.09}$ $0.61^{+0.13}_{-0.13}$ | $3.05(>1.84)$ $0.25^{+0.06}_{-0.06}$ | $2.08^{+0.35}_{-0.40}$ ... | $1.92^{+0.37}_{-0.21}$ ... | $0.21^{+0.82}_{-0.15}$ $6.39^{+0.01}_{-0.01}$ | $1.09^{+3.51}_{-0.89}$ $1.18^{+0.14}_{-0.14}$ | ... $7.06$ (fixed) | $0.91(>0.35)$ $0.10(<0.20)$ | 55.7/63.0 |
| [1]Mrk 3 | $1.01^{+0.02}_{-0.02}$ ... | $1.45^{+0.10}_{-0.08}$ ... | $0.87^{+0.05}_{-0.05}$ ... | $1.74^{+0.04}_{-0.04}$ ... | $0.57^{+0.10}_{-0.09}$ $6.40^{+0.01}_{-0.01}$ | $3.12^{+0.76}_{-0.59}$ $5.06^{+0.22}_{-0.22}$ | ... $7.06$ (fixed) | $2.00(>1.84)$ $0.58^{+0.16}_{-0.16}$ | 890.1/701.0 |
| NGC 1194 | $0.93^{+0.09}_{-0.08}$ ... | $1.35^{+0.50}_{-0.43}$ ... | $1.15^{+0.37}_{-0.28}$ ... | $1.71^{+0.16}_{-0.18}$ ... | $0.14^{+0.17}_{-0.07}$ $6.42^{+0.01}_{-0.01}$ | $2.28^{+2.04}_{-1.23}$ $1.32^{+0.19}_{-0.19}$ | ... $7.06$ (fixed) | $2.00(>1.12)$ $0.25^{+0.15}_{-0.15}$ | 62.0/53.0 |
| NGC 3393 | $1.05^{+0.20}_{-0.16}$ $0.16^{+0.04}_{-0.04}$ | $0.89^{+0.39}_{-0.29}$ $1.54^{+1.62}_{-0.66}$ | $6.52^{+2.98}_{-4.50}$ $0.79^{+0.09}_{-0.08}$ | $2.49(>1.71)$ $0.42^{+0.16}_{-0.13}$ | $5.08^{+1.00}_{-4.86}$ $6.42^{+0.01}_{-0.02}$ | $0.09^{+1.39}_{-0.03}$ $0.75^{+0.14}_{-0.14}$ | ... $7.06$ (fixed) | $0.04^{+0.06}_{-0.01}$ $0.22(<0.35)$ | 32.1/35.0 |
| [1]NGC 4945 | $1.04^{+0.03}_{-0.03}$ ... | $1.50^{+0.07}_{-0.07}$ ... | $4.51^{+0.53}_{-0.51}$ ... | $1.66^{+0.08}_{-0.07}$ ... | $1.89^{+0.72}_{-0.50}$ $6.40^{+0.01}_{-0.01}$ | $1.92^{+0.58}_{-0.46}$ $3.23^{+0.17}_{-0.17}$ | ... $7.06$ (fixed) | $0.25^{+0.03}_{-0.03}$ $0.44^{+0.11}_{-0.11}$ | 496.0/402.0 |
| NGC 5728 | $0.99^{+0.09}_{-0.09}$ $0.29^{+0.35}_{-0.06}$ | $0.85^{+0.16}_{-0.15}$ $0.58^{+0.36}_{-0.26}$ | $1.30^{+0.18}_{-0.22}$ ... | $1.63^{+0.14}_{-0.13}$ ... | $0.57^{+0.41}_{-0.26}$ $6.39^{+0.01}_{-0.01}$ | $0.72^{+0.83}_{-0.38}$ $2.26^{+0.23}_{-0.23}$ | ... $7.06$ (fixed) | $0.13^{+1.08}_{-0.09}$ $0.16^{+0.29}_{-0.17}$ | 72.0/61.0 |
| NGC 6552 | $0.90^{+0.09}_{-0.08}$ $0.59^{+0.14}_{-0.16}$ | $1.20^{+0.54}_{-0.62}$ $0.10^{+0.03}_{-0.03}$ | $0.75^{+0.28}_{-0.17}$ ... | $1.50(<1.72)$ ... | $0.03^{+0.08}_{-0.01}$ $6.40^{+0.01}_{-0.01}$ | $2.35^{+1.21}_{-1.69}$ $1.38^{+0.13}_{-0.12}$ | ... $7.06$ (fixed) | $2.00(>0.76)$ $0.13^{+0.07}_{-0.07}$ | 80.5/69.0 |
| NGC 7130 | $1.00^{+0.23}_{-0.14}$ $0.89^{+0.10}_{-0.13}$ | $0.73^{+0.55}_{-0.41}$ $0.42^{+0.20}_{-0.15}$ | $2.48^{+0.73}_{-0.56}$ ... | $2.46(>2.21)$ ... | $2.16^{+1.55}_{-1.29}$ $6.42^{+0.06}_{-0.03}$ | $0.36^{+0.44}_{-0.16}$ $0.44^{+0.16}_{-0.15}$ | ... $7.06$ (fixed) | $0.01^{+0.04}_{-0.01}$ $0.00(<0.01)$ | 23.6/23.0 |
| NGC 7582 | $1.02^{+0.07}_{-0.17}$ $0.78^{+0.21}_{-0.19}$ | $0.60^{+0.20}_{-0.21}$ $0.52^{+0.15}_{-0.19}$ | $0.91^{+0.25}_{-0.22}$ ... | $1.76^{+0.09}_{-0.09}$ ... | $0.45^{+0.28}_{-0.20}$ $6.40^{+0.01}_{-0.01}$ | $2.91^{+2.33}_{-1.21}$ $2.22^{+0.34}_{-0.34}$ | $0.06^{+0.05}_{-0.05}$ $7.06$ (fixed) | $1.71(>0.99)$ $0.01(<0.23)$ | 99.3/70.0 |

Note. (1) Galaxy name. (2) Cross-calibration constant (const1) of the BIXIS relative to the FIXISs. (3) Time variability constant (const2) of the direct component between the *Suzaku* and *Swift*. (4) Hydrogen column density of the direct component in units of $10^{24}$ cm$^{-2}$. (5) Photon index. (6) Power law normalization of the direct component in units of $10^{-2}$ photons keV$^{-1}$ cm$^{-2}$ s$^{-1}$. (7) Scattering fraction in units of percent. (8) Hydrogen column density of the reflection component in units of $10^{24}$ cm$^{-2}$. (9) Relative reflection strength in units of $\Omega/2\pi$ where $\Omega$ is the solid angle of the reflector. (10) Temperature of the apec1 model in units of keV. (11) Normalization of the apec1 model in units of $10^{-18}/4\pi [D_A(1+z)]^2 \int n_e n_H dV$, where $D_A$ is the angular diameter distance to the source (cm), $n_e$ and $n_H$ are the electron and hydrogen densities (cm$^{-3}$). (12) Temperature of the apec2 model in units of keV. (13) Normalization of the apec2 model in units of $10^{-18}/4\pi [D_A(1+z)]^2 \int n_e n_H dV$, where $D_A$ is the angular diameter distance to the source (cm), $n_e$ and $n_H$ are the electron and hydrogen densities (cm$^{-3}$). (14) Central energy of the Fe K$\alpha$ emission line in units of keV. (15) Normalization of the Fe K$\alpha$ emission line in units of $10^{-5}$ photons cm$^{-2}$ s$^{-1}$. (16) Central energy of the Fe K$\beta$ emission line in units of keV. (17) Normalization of the Fe K$\beta$ emission line in units of $10^{-5}$ photons cm$^{-2}$ s$^{-1}$. See Section 4.5 and 4.8 for details.

[1]For these targets, we also consider emission lines other than Fe.

TABLE 5
Best fit parameters with the Baseline2 model

| (1) Galaxy Name | (2) $N_{\rm BIXIS}$ (10) $k_B T_1$ | (3) $N_{\rm Suzaku}$ (11) $N_{\rm Apec1}$ | (4) $N_{\rm H}^{\rm Dir}$ (12) $k_B T_2$ | (5) $\Gamma$ (13) $N_{\rm Apec2}$ | (6) $N_{\rm Dir}^{\rm FeK\alpha}$ (14) $E_{\rm FeK\alpha}$ | (7) $f_{\rm scat}$ (15) $I_{\rm FeK\alpha}$ | (8) $N_{\rm H}^{\rm Ref}$ (16) $E_{\rm FeK\beta}$ | (9) $R_{\rm Ref}$ (17) $I_{\rm FeK\beta}$ | $\chi^2$/dof |
|---|---|---|---|---|---|---|---|---|---|
| CGCG 420-015 | $1.04^{+0.05}_{-0.05}$ $0.18^{+0.05}_{-0.04}$ | $0.60^{+0.18}_{-0.13}$ $0.78^{+0.68}_{-0.37}$ | $0.91^{+0.14}_{-0.12}$ $0.99^{+0.10}_{-0.15}$ | $2.38(>2.21)$ $0.37^{+0.10}_{-0.09}$ | $4.36^{+3.14}_{-2.21}$ $6.39^{+0.01}_{-0.01}$ | $0.14^{+0.12}_{-0.06}$ $1.07^{+0.15}_{-0.15}$ | $0.09^{+0.05}_{-0.04}$ 7.06 (fixed) | $0.26^{+0.18}_{-0.10}$ $0.06(<0.19)$ | 154.6/154.0 |
| ESO 137-G034 | $0.91^{+0.09}_{-0.14}$ $0.66^{+0.14}_{-0.09}$ | $0.37^{+0.20}_{-0.14}$ $0.60^{+0.15}_{-0.15}$ | $1.48^{+0.41}_{-0.39}$ ... | $2.25(>1.85)$ ... | $4.27^{+8.46}_{-3.46}$ $6.39^{+0.01}_{-0.01}$ | $0.17^{+0.60}_{-0.12}$ $0.88^{+0.15}_{-0.12}$ | ... 7.06 (fixed) | $0.05^{+0.04}_{-0.02}$ $0.11^{+0.09}_{-0.09}$ | 52.6/64.0 |
| ESO 323-G032 | $1.05^{+0.14}_{-0.13}$ $0.74^{+0.81}_{-0.46}$ | $1.10^{+1.51}_{-0.64}$ $0.05^{+0.05}_{-0.02}$ | $1.99^{+0.53}_{-0.69}$ ... | $2.16(>1.61)$ ... | $2.47^{+12.2}_{-2.33}$ $6.39^{+0.02}_{-0.02}$ | $0.08^{+1.17}_{-0.07}$ $0.82^{+0.12}_{-0.12}$ | ... 7.06 (fixed) | $0.08^{+0.41}_{-0.05}$ $0.00(<0.07)$ | 41.4/37.0 |
| ESO 565-G019 | $0.89^{+0.09}_{-0.09}$ $0.61^{+0.13}_{-0.13}$ | $2.97(>1.84)$ $0.25^{+0.06}_{-0.06}$ | $2.07^{+0.33}_{-0.39}$ ... | $1.99^{+0.38}_{-0.39}$ ... | $1.31^{+6.30}_{-1.15}$ $6.39^{+0.01}_{-0.01}$ | $0.18^{+0.86}_{-0.14}$ $1.18^{+0.14}_{-0.14}$ | ... 7.06 (fixed) | $0.17^{+0.34}_{-0.11}$ $0.10(<0.20)$ | 54.8/63.0 |
| [1]Mrk 3 | $1.01^{+0.02}_{-0.02}$ ... | $1.55^{+0.22}_{-0.17}$ ... | $0.86^{+0.06}_{-0.06}$ ... | $1.74^{+0.05}_{-0.05}$ ... | $1.02^{+0.36}_{-0.28}$ $6.40^{+0.01}_{-0.01}$ | $1.69^{+0.66}_{-0.47}$ $5.08^{+0.22}_{-0.22}$ | ... 7.06 (fixed) | $1.16^{+0.31}_{-0.22}$ $0.60^{+0.16}_{-0.16}$ | 884.9/701.0 |
| NGC 1194 | $0.92^{+0.09}_{-0.08}$ ... | $1.94^{+1.11}_{-0.93}$ ... | $1.10^{+0.41}_{-0.29}$ ... | $1.67^{+0.14}_{-0.16}$ ... | $0.16^{+0.72}_{-0.08}$ $6.42^{+0.01}_{-0.01}$ | $2.15^{+1.75}_{-1.38}$ $1.35^{+0.19}_{-0.19}$ | ... 7.06 (fixed) | $2.00(>0.46)$ $0.27^{+0.15}_{-0.15}$ | 60.9/53.0 |
| NGC 3393 | $1.05^{+0.20}_{-0.17}$ $0.16^{+0.04}_{-0.04}$ | $0.92^{+0.40}_{-0.30}$ $1.54^{+2.15}_{-0.67}$ | $4.57^{+1.88}_{-2.63}$ $0.79^{+0.09}_{-0.08}$ | $2.50(>1.77)$ $0.42^{+0.16}_{-0.13}$ | $119^{+370}_{-118}$ $6.41^{+0.02}_{-0.02}$ | $0.01^{+0.29}_{-0.01}$ $0.74^{+0.14}_{-0.14}$ | ... 7.06 (fixed) | $0.02^{+0.02}_{-0.01}$ $0.20(<0.34)$ | 32.4/35.0 |
| [1]NGC 4945 | $1.04^{+0.03}_{-0.03}$ ... | $1.47^{+0.07}_{-0.06}$ ... | $4.40^{+0.53}_{-0.50}$ ... | $1.99^{+0.13}_{-0.12}$ ... | $143^{+196}_{-80.3}$ $6.40^{+0.01}_{-0.01}$ | $0.03^{+0.04}_{-0.02}$ $3.20^{+0.17}_{-0.17}$ | ... 7.06 (fixed) | $0.01^{+0.01}_{-0.01}$ $0.41^{+0.11}_{-0.11}$ | 525.6/402.0 |
| NGC 5728 | $0.98^{+0.09}_{-0.09}$ $0.27^{+0.15}_{-0.07}$ | $0.85^{+0.16}_{-0.15}$ $0.54^{+0.37}_{-0.26}$ | $1.37^{+0.19}_{-0.25}$ ... | $1.73^{+0.15}_{-0.14}$ ... | $1.85^{+2.00}_{-1.13}$ $6.39^{+0.01}_{-0.01}$ | $0.29^{+0.43}_{-0.14}$ $2.28^{+0.23}_{-0.23}$ | $0.22^{+0.27}_{-0.21}$ 7.06 (fixed) | $0.15(<0.66)$ $0.18^{+0.17}_{-0.17}$ | 69.8/61.0 |
| NGC 6552 | $0.91^{+0.08}_{-0.14}$ $0.59^{+0.14}_{-0.16}$ | $1.28^{+0.55}_{-0.68}$ $0.10^{+0.03}_{-0.03}$ | $0.74^{+0.27}_{-0.17}$ ... | $1.50(<1.73)$ ... | $0.06^{+0.17}_{-0.02}$ $6.40^{+0.01}_{-0.01}$ | $1.39^{+1.14}_{-1.14}$ $1.38^{+0.13}_{-0.12}$ | ... 7.06 (fixed) | $1.22(>0.48)$ $0.13^{+0.07}_{-0.07}$ | 80.5/69.0 |
| NGC 7130 | $0.99^{+0.23}_{-0.18}$ $0.90^{+0.10}_{-0.13}$ | $0.75^{+0.58}_{-0.45}$ $0.42^{+0.21}_{-0.15}$ | $2.43^{+1.17}_{-0.68}$ ... | $2.50(>2.23)$ ... | $13.9^{+45.0}_{-10.8}$ $6.43^{+0.06}_{-0.06}$ | $0.06^{+0.12}_{-0.03}$ $0.43^{+0.16}_{-0.15}$ | ... 7.06 (fixed) | $0.01^{+0.01}_{-0.01}$ $0.00(<0.01)$ | 23.4/23.0 |
| NGC 7582 | $1.03^{+0.07}_{-0.07}$ $0.78^{+0.18}_{-0.22}$ | $0.60^{+0.20}_{-0.21}$ $0.51^{+0.18}_{-0.17}$ | $0.94^{+0.26}_{-0.24}$ ... | $1.81^{+0.10}_{-0.10}$ ... | $1.05^{+0.95}_{-0.52}$ $6.40^{+0.01}_{-0.01}$ | $1.40^{+1.53}_{-0.99}$ $2.21^{+0.35}_{-0.34}$ | $0.08^{+0.06}_{-0.05}$ 7.06 (fixed) | $0.84^{+1.09}_{-0.37}$ $0.01(<0.28)$ | 99.4/70.0 |

Note. (1) Galaxy name. (2) Cross-calibration constant (const1) of the BIXIS relative to the FIXISs. (3) Time variability constant (const2) of the BIXIS between the *Suzaku* and *Swift*. (4) Hydrogen column density of the direct component in units of $10^{24}$ cm$^{-2}$. (5) Photon index. (6) Power law normalization of the direct component in units of $10^{-2}$ photons keV$^{-1}$ cm$^{-2}$ s$^{-1}$. (7) Scattering fraction in units of percent. (8) Hydrogen column density of the reflection component in units of $10^{24}$ cm$^{-2}$. (9) Relative reflection strength in units of $\Omega/2\pi$ where $\Omega$ is the solid angle of the reflector. (10) Temperature of the apec1 model in units of keV. (11) Normalization of the apec1 model in units of $10^{-18}/4\pi[D_A(1+z)]^2 \int n_e n_H dV$, where $D_A$ is the angular diameter distance to the source (cm), $n_e$ and $n_H$ are the electron and hydrogen densities (cm$^{-3}$). (12) Temperature of the apec2 model in units of keV. (13) Normalization of the apec2 model in units of $10^{-18}/4\pi[D_A(1+z)]^2 \int n_e n_H dV$, where $D_A$ is the angular diameter distance to the source (cm), $n_e$ and $n_H$ are the electron and hydrogen densities (cm$^{-3}$). (14) Central energy of the Fe K$\alpha$ emission line in units of keV. (15) Normalization of the Fe K$\alpha$ emission line in units of $10^{-5}$ photons cm$^{-2}$ s$^{-1}$. (16) Central energy of the Fe K$\beta$ emission line in units of keV. (17) Normalization of the Fe K$\beta$ emission line in units of $10^{-5}$ photons cm$^{-2}$ s$^{-1}$. See Section 4.5 and 4.8 for details.
[1]For these targets, we also consider emission lines other than Fe.

TABLE 6
BEST FIT PARAMETERS WITH THE IKEDA1 MODEL

| (1) Galaxy Name | (2) $N_{\rm BIXIS}$ (9) $N_{\rm Dir}$ | (3) $N_{\rm Suzaku}$ (10) $f_{\rm scat}$ | (4) $N_{\rm H}^{\rm Equ}$ (11) $N_{\rm FeK\alpha}$ | (5) $N_{\rm H}^{\rm LOS}$ (12) $E_{\rm FeK\beta}$ | (6) $\theta_{\rm open}$ (13) $I_{\rm FeK\beta}$ | (7) $\theta_{\rm incl}$ (14) $EW_{\rm FeK\alpha}$ | (8) $\Gamma$ (15) $EW_{\rm FeK\beta}$ | $\chi^2/$dof |
|---|---|---|---|---|---|---|---|---|
| CGCG 420-015 | $1.05^{+0.05}_{-0.05}$ $2.31^{+1.61}_{-0.72}$ | $0.39^{+0.16}_{-0.12}$ $0.23^{+0.10}_{-0.09}$ | $0.91^{+0.13}_{-0.12}$ $0.26^{+0.06}_{-0.04}$ | $0.79^{+0.11}_{-0.10}$ $7.06$ (fixed) | $19.9^{+25.1}_{-1.91}$ $0.20^{+0.12}_{-0.12}$ | $23.2^{+19.0}_{-0.82}$ $0.27^{+0.16}_{-0.08}$ | $2.30^{+0.15}_{-0.15}$ $0.06^{+0.04}_{-0.04}$ | 148.0/145.0 |
| ESO 137-G034 | $0.93^{+0.10}_{-0.09}$ $0.98^{+1.74}_{-0.74}$ | $0.23(<0.55)$ $0.70^{+1.45}_{-0.41}$ | $1.49^{+0.63}_{-0.62}$ $1.08^{+0.84}_{-0.39}$ | $1.45^{+0.61}_{-0.60}$ $7.06$ (fixed) | $70.0(>48.2)$ $0.17^{+0.09}_{-0.09}$ | $74.8(>66.1)$ $1.04^{+0.70}_{-0.72}$ | $1.93^{+0.26}_{-0.32}$ $0.24^{+0.12}_{-0.14}$ | 50.1/57.0 |
| ESO 323-G032 | $1.03^{+0.14}_{-0.13}$ $1.27^{+25.0}_{-1.02}$ | $1.45(>0.20)$ $0.21^{+0.86}_{-0.19}$ | $2.49^{+4.60}_{-0.73}$ $0.90^{+0.86}_{-0.33}$ | $2.36^{+4.37}_{-0.69}$ $7.06$ (fixed) | $67.9(>10.0)$ $0.03(<0.12)$ | $71.3^{+4.31}_{-8.61}$ $1.29^{+28.7}_{-0.99}$ | $2.08^{+0.42}_{-0.38}$ $0.06(<0.22)$ | 35.6/35.0 |
| ESO 565-G019 | $0.88^{+0.09}_{-0.08}$ $1.01^{+0.65}_{-0.39}$ | $4.83(>2.95)$ $0.27^{+0.13}_{-0.11}$ | $7.78(>3.42)$ $1.13^{+0.10}_{-0.16}$ | $2.05^{+2.64}_{-1.15}$ $7.06$ (fixed) | $10.0(<18.7)$ $0.13^{+0.10}_{-0.10}$ | $10.2^{+2.54}_{-0.03}$ $1.14^{+0.48}_{-0.37}$ | $2.10^{+0.23}_{-0.19}$ $0.13^{+0.10}_{-0.07}$ | 50.9/58.0 |
| [1]Mrk 3 | $1.00^{+0.02}_{-0.02}$ $5.69^{+0.29}_{-0.35}$ | $1.11^{+0.06}_{-0.06}$ $0.35^{+0.03}_{-0.03}$ | $10.0(>9.86)$ $0.53^{+0.03}_{-0.03}$ | $1.54(>1.52)$ $7.06$ (fixed) | $14.5^{+0.01}_{-0.01}$ $0.58^{+0.16}_{-0.16}$ | $14.6^{+0.01}_{-0.01}$ $0.39^{+0.01}_{-0.02}$ | $2.10^{+0.02}_{-0.03}$ $0.05^{+0.03}_{-0.01}$ | 1002.4/658.0 |
| NGC 1194 | $0.93^{+0.08}_{-0.08}$ $0.69^{+0.20}_{-0.22}$ | $1.25^{+0.52}_{-0.39}$ $0.29^{+0.10}_{-0.09}$ | $6.02^{+1.17}_{-0.50}$ $0.69^{+0.20}_{-0.15}$ | $0.94^{+0.18}_{-0.08}$ $7.06$ (fixed) | $18.9^{+3.10}_{-2.50}$ $0.28^{+0.15}_{-0.15}$ | $19.0^{+6.89}_{-0.11}$ $0.52^{+0.11}_{-0.14}$ | $1.93^{+0.07}_{-0.08}$ $0.14^{+0.08}_{-0.08}$ | 72.9/52.0 |
| NGC 3393 | $1.03^{+0.16}_{-0.14}$ $17.0^{+86.6}_{-16.1}$ | $0.24(>0.20)$ $0.03^{+1.30}_{-0.02}$ | $5.04^{+3.55}_{-2.53}$ $0.85^{+0.53}_{-0.26}$ | $4.98^{+3.51}_{-2.51}$ $7.06$ (fixed) | $69.8(>10.0)$ $0.19^{+0.13}_{-0.13}$ | $79.2^{+1.52}_{-50.6}$ $2.10^{+45.8}_{-2.00}$ | $2.29(>1.71)$ $0.74^{+0.52}_{-0.51}$ | 23.3/28.0 |
| [1]NGC 4945 | $1.05^{+0.03}_{-0.03}$ $1.16^{+0.09}_{-0.02}$ | $4.45^{+3.97}_{-0.13}$ $2.94^{+0.13}_{-0.18}$ | $3.39^{+0.26}_{-0.24}$ $0.84^{+0.08}_{-0.07}$ | $2.93^{+0.22}_{-0.21}$ $7.06$ (fixed) | $11.7^{+0.56}_{-0.28}$ $0.44^{+0.12}_{-0.11}$ | $15.0^{+0.76}_{-1.37}$ $0.64^{+0.06}_{-0.05}$ | $1.50(<1.52)$ $0.13^{+0.04}_{-0.02}$ | 491.4/402.0 |
| NGC 5728 | $0.99^{+0.09}_{-0.09}$ $0.69^{+0.44}_{-0.28}$ | $0.71^{+0.30}_{-0.25}$ $0.70^{+0.34}_{-0.26}$ | $1.34^{+0.19}_{-0.19}$ $0.68^{+0.22}_{-0.14}$ | $1.24^{+0.18}_{-0.18}$ $7.06$ (fixed) | $19.0^{+10.8}_{-1.36}$ $0.33^{+0.16}_{-0.16}$ | $19.1^{+54.5}_{-6.89}$ $0.87^{+0.45}_{-0.27}$ | $1.65^{+0.13}_{-0.13}$ $0.16^{+0.07}_{0.08}$ | 64.0/58.0 |
| NGC 6552 | $0.89^{+0.08}_{-0.08}$ $0.18^{+0.14}_{-0.10}$ | $0.76^{+0.61}_{-0.23}$ $0.60^{+0.18}_{-0.20}$ | $7.25(>2.13)$ $1.78^{+2.14}_{-0.49}$ | $1.14(>0.34)$ $7.06$ (fixed) | $19.0^{+10.8}_{-1.36}$ $0.14^{+0.08}_{-0.08}$ | $19.1^{+10.8}_{-1.38}$ $1.19^{+2.14}_{-0.45}$ | $1.76(<1.91)$ $0.15^{+0.08}_{-0.08}$ | 90.4/65.0 |
| NGC 7130 | $0.99^{+0.20}_{-0.17}$ $4.19^{+183}_{-2.32}$ | $0.63(<1.71)$ $0.17^{+0.69}_{-0.16}$ | $2.25^{+5.66}_{-0.46}$ $1.30^{+1.14}_{-0.94}$ | $2.25^{+5.66}_{-0.46}$ $7.06$ (fixed) | $70.0(>10.0)$ $0.00(<0.01)$ | $89.0(>62.6)$ $1.28^{+100}_{-1.14}$ | $2.35(>2.01)$ $0.00(<0.86)$ | 19.1/17.0 |
| NGC 7582 | $1.03^{+0.07}_{-0.07}$ $0.90^{+0.36}_{-0.13}$ | $0.57^{+0.18}_{-0.15}$ $1.24^{+0.31}_{-0.17}$ | $2.36^{+2.23}_{-0.51}$ $0.50^{+0.08}_{-0.08}$ | $0.71^{+0.67}_{-0.15}$ $7.06$ (fixed) | $34.5^{+2.97}_{-6.75}$ $0.12(<0.39)$ | $34.7^{+29.9}_{-1.07}$ $0.45^{+0.19}_{-0.12}$ | $1.80^{+0.09}_{-0.10}$ $0.03(<0.09)$ | 94.6/66.0 |

Note. (1) Galaxy name. (2) Cross-calibration constant (const1) of the FIXISs. (3) Time variability constant (const2) of the direct component between the *Suzaku* and *Swift*. (4) Hydrogen column density along the equatorial plane in units of $10^{24}$ cm$^{-2}$. (5) Hydrogen column density along the line-of-sight in units of $10^{24}$ cm$^{-2}$. (6) Half-opening angle of the torus in units of degree. (7) Inclination angle of the torus in units of degree. (8) Photon index. (9) Power law normalization of the direct component in units of $10^{-2}$ photons keV$^{-1}$ cm$^{-2}$ s$^{-1}$. (10) Scattering fraction in units of percent. (11) Relative normalization of the Fe K$\alpha$ emission line to the reflection component. (12) Energy of the Fe K$\beta$ emission line in units of keV. (13) Normalization of the Fe K$\beta$ emission line in units of $10^{-5}$ total photons cm$^{-2}$ s$^{-1}$. (14) Equivalent width of the Fe K$\alpha$ emission line with respect to the whole continuum in units of keV. (15) Equivalent width of the Fe K$\beta$ emission line with respect to the whole continuum in units of keV. See Section 4.5 and 4.8 for details.
[1]For these targets, we also consider emission lines other than Fe.

TABLE 7
BEST FIT PARAMETERS WITH THE IKEDA2 MODEL

| (1) Galaxy Name | (2) $N_{\rm BIXIS}$ (9) $N_{\rm Dir}$ | (3) $N_{\rm Suzaku}$ (10) $f_{\rm scat}$ | (4) $N_{\rm H}^{\rm Dir}$ (11) $N_{\rm FeK\alpha}$ | (5) $N_{\rm H}^{\rm Equ}$ (12) $E_{\rm FeK\beta}$ | (6) $\theta_{\rm open}$ (13) $I_{\rm FeK\beta}$ | (7) $\theta_{\rm incl}$ (14) $EW_{\rm FeK\alpha}$ | (8) $\Gamma$ (15) $EW_{\rm FeK\beta}$ | $\chi^2/{\rm dof}$ |
|---|---|---|---|---|---|---|---|---|
| CGCG 420-015 | $1.05^{+0.05}_{-0.05}$ $4.63^{+7.56}_{-2.88}$ | $0.53(<0.77)$ $0.12^{+0.18}_{-0.08}$ | $1.23^{+1.65}_{-0.27}$ $0.29^{+0.12}_{-0.15}$ | $0.66^{+0.29}_{-0.24}$ 7.06 (fixed) | $64.7^{+1.00}_{-3.16}$ $0.18^{+0.22}_{-0.13}$ | 70.0 (fixed) $0.26^{+0.05}_{-0.10}$ | $2.34(>2.08)$ $0.05^{+0.12}_{-0.01}$ | 144.8/145.0 |
| ESO 137-G034 | $0.93^{+0.09}_{-0.08}$ $1.89^{+5.84}_{-1.38}$ | $0.33(<0.56)$ $0.37^{+0.83}_{-0.27}$ | $1.30^{+0.37}_{-0.34}$ $1.36^{+1.96}_{-0.49}$ | $10.0(>2.32)$ 7.06 (fixed) | $65.9^{+0.90}_{-0.40}$ $0.15^{+0.09}_{-0.09}$ | 70.0 (fixed) $1.03^{+1.48}_{-0.66}$ | $2.11^{+0.34}_{-0.32}$ $0.21^{+0.11}_{-0.13}$ | 47.2/57.0 |
| ESO 323-G032 | $1.03^{+0.14}_{-0.13}$ $2.90^{+9.98}_{-2.58}$ | $1.25(>0.20)$ $0.09^{+0.57}_{-0.07}$ | $2.24^{+5.29}_{-0.74}$ $0.90^{+0.44}_{-0.34}$ | $5.01(>1.51)$ 7.06 (fixed) | $66.0^{+3.35}_{-1.75}$ $0.03(<0.11)$ | 70.0 (fixed) $1.27^{+0.62}_{-0.48}$ | $2.30(>2.02)$ $0.05(<0.19)$ | 35.0/35.0 |
| ESO 565-G019 | $0.88^{+0.09}_{-0.08}$ $0.87^{+1.00}_{-0.56}$ | $3.96(>2.42)$ $0.29^{+0.39}_{-0.16}$ | $2.17^{+0.50}_{-0.30}$ $1.20^{+0.33}_{-0.40}$ | $5.02(>1.11)$ 7.06 (fixed) | $69.0(>65.8)$ $0.14^{+0.05}_{-0.10}$ | 70.0 (fixed) $1.16^{+0.32}_{-0.39}$ | $1.95^{+0.50}_{-0.23}$ $0.15^{+0.05}_{-0.11}$ | 50.7/58.0 |
| [1]Mrk 3 | $1.00^{+0.02}_{-0.02}$ $1.40^{+0.28}_{-0.10}$ | $1.36^{+0.07}_{-0.07}$ $1.34^{+0.21}_{-0.23}$ | $1.11^{+0.09}_{-0.04}$ $0.65^{+0.05}_{-0.12}$ | $0.31(<0.39)$ 7.06 (fixed) | 50.0 (fixed) $0.71^{+0.16}_{-0.16}$ | 51.0 (fixed) $0.42^{+0.03}_{-0.08}$ | $1.60^{+0.05}_{-0.02}$ $0.07^{+0.02}_{-0.02}$ | 766.7/659.0 |
| NGC 1194 | $0.93^{+0.09}_{-0.08}$ $0.58^{+1.14}_{-0.21}$ | $1.38^{+0.23}_{-0.34}$ $0.37^{+0.41}_{-0.25}$ | $1.55^{+1.48}_{-0.51}$ $0.62^{+0.58}_{-0.27}$ | $0.50^{+1.00}_{-0.32}$ 7.06 (fixed) | $69.3(>65.1)$ $0.30^{+0.16}_{-0.14}$ | 70.0 (fixed) $0.57^{+0.55}_{-0.24}$ | $1.63^{+0.25}_{-0.12}$ $0.16^{+0.08}_{-0.08}$ | 72.2/52.0 |
| NGC 3393 | $1.07^{+0.17}_{-0.16}$ $4.51^{+19.5}_{-2.43}$ | $0.84(>0.20)$ $0.10^{+0.55}_{-0.02}$ | $2.78(>1.48)$ $1.18^{+19.4}_{-0.67}$ | $10.0(>1.40)$ 7.06 (fixed) | $61.0^{+4.21}_{-5.66}$ $0.16(<0.33)$ | 70.0 (fixed) $2.13^{+33.9}_{-0.92}$ | $2.11(>1.72)$ $0.54(<1.10)$ | 24.3/28.0 |
| [1]NGC 4945 | $1.05^{+0.03}_{-0.03}$ $7.31^{+1.75}_{-1.50}$ | $2.33^{+0.38}_{-0.26}$ $0.52^{+0.11}_{-0.09}$ | $3.45^{+0.38}_{-0.32}$ $0.57^{+0.08}_{-0.05}$ | $10.0(>9.08)$ 7.06 (fixed) | $65.3^{+0.13}_{-0.11}$ $0.48^{+0.11}_{-0.09}$ | 70.0 (fixed) $0.63^{+0.09}_{-0.05}$ | $1.62^{+0.04}_{-0.04}$ $0.14^{+0.04}_{-0.03}$ | 505.2/402.0 |
| NGC 5728 | $0.99^{+0.09}_{-0.09}$ $1.44^{+1.12}_{-0.66}$ | $0.79^{+0.29}_{-0.25}$ $0.34^{+0.25}_{-0.14}$ | $1.69^{+1.45}_{-0.53}$ $0.64^{+0.60}_{-0.19}$ | $1.24^{+2.82}_{-0.32}$ 7.06 (fixed) | $60.3^{+5.45}_{-12.3}$ $0.36^{+0.17}_{-0.18}$ | 70.0 (fixed) $0.87^{+0.80}_{-0.26}$ | $1.69^{+0.14}_{-0.09}$ $0.16^{+0.08}_{-0.08}$ | 64.0/65.0 |
| NGC 6552 | $0.90^{+0.08}_{-0.08}$ $0.33^{+0.22}_{-0.20}$ | $1.27^{+1.65}_{-0.64}$ $0.32^{+0.67}_{-0.16}$ | $1.74^{+1.56}_{-0.99}$ $1.12^{+2.15}_{-0.33}$ | $0.50(<0.88)$ 7.06 (fixed) | $67.1(>53.9)$ $0.15^{+0.09}_{-0.09}$ | 70.0 (fixed) $1.24^{+2.37}_{-0.37}$ | $1.62(<1.76)$ $0.16^{+0.09}_{-0.10}$ | 84.1/65.0 |
| NGC 7130 | $0.93^{+0.25}_{-0.13}$ $3.59^{+11.0}_{-2.96}$ | $0.53(<1.50)$ $0.24^{+0.65}_{-0.14}$ | $1.69^{+2.82}_{-0.54}$ $7.12^{+41.9}_{-6.50}$ | $5.02(>1.99)$ 7.06 (fixed) | $10.0(<46.5)$ $0.00(<0.01)$ | 70.0 (fixed) $1.34^{+8.07}_{-1.22}$ | $2.50(>2.27)$ $0.02(<0.03)$ | 18.6/17.0 |
| NGC 7582 | $1.02^{+0.07}_{-0.07}$ $4.92^{+4.92}_{-2.69}$ | $0.20(<0.76)$ $0.23^{+0.26}_{-0.07}$ | $3.23^{+1.33}_{-1.78}$ $0.30^{+0.10}_{-0.06}$ | $1.46^{+0.27}_{-0.72}$ 7.06 (fixed) | $66.5^{+2.57}_{-0.61}$ $0.26(<0.52)$ | 70.0 (fixed) $0.44^{+0.15}_{-0.09}$ | $1.88^{+0.11}_{-0.12}$ $0.07^{+0.11}_{-0.13}$ | 93.1/66.0 |

Note. (1) Galaxy name. (2) Cross-calibration constant (**const1**) of the BIXIS relative to the FIXISs. (3) Time variability constant (**const2**) of the direct component between the *Suzaku* and *Swift*. (4) Hydrogen column density of the direct component in units of $10^{24}$ cm$^{-2}$. (5) Hydrogen column density along the line-of-sight in units of $10^{24}$ cm$^{-2}$. (6) Half-opening angle of the torus in units of degree. (7) Inclination angle of the torus in units of degree. (8) Photon index. (9) Power law normalization of the direct component in units of $10^{-2}$ photons keV$^{-1}$ cm$^{-2}$ s$^{-1}$. (10) Scattering fraction in units of percent. (11) Relative normalization of the Fe K$\alpha$ emission line to the reflection component. (12) Energy of the Fe K$\beta$ emission line in units of keV. (13) Normalization of the Fe K$\beta$ emission line in units of $10^{-5}$ total photons cm$^{-2}$ s$^{-1}$. (14) Equivalent width of the Fe K$\alpha$ emission line with respect to the whole continuum in units of keV. (15) Equivalent width of the Fe K$\beta$ emission line with respect to the whole continuum in units of keV. See Section 4.5 and 4.8 for details.
[1]For these targets, we also consider emission lines other than Fe.

TABLE 8
Fluxes and Luminosities with the Baseline2 model

| (1) Galaxy Name | (2) log $F^{\text{Suzaku}}_{0.5-2}$ | (3) log $F^{\text{Suzaku}}_{2-10}$ | (4) log $F^{\text{Suzaku}}_{10-50}$ | (5) log $F^{\text{Swift}}_{10-50}$ | (6) log $L^{\text{Suzaku}}_{0.5-2}$ | (7) log $L^{\text{Suzaku}}_{2-10}$ | (8) log $L^{\text{Suzaku}}_{10-50}$ | (9) log $L^{\text{Swift}}_{10-50}$ | (10) log $L_{\text{Fe}K\alpha}$ | (11) log $\lambda_{\text{Edd}}$ |
|---|---|---|---|---|---|---|---|---|---|---|
| CGCG 420−015 | -12.68 | -11.75 | -10.88 | -10.79 | 44.04 | 43.85 | 43.57 | 43.78 | 41.32 | - 1.04 |
| ESO 137−G034 | -12.85 | -12.26 | -11.20 | -10.90 | 42.80 | 42.70 | 42.51 | 42.93 | 40.20 | - 1.69 |
| ESO 323−G032 | -13.33 | -12.40 | -11.06 | -11.17 | 43.53 | 43.48 | 43.34 | 43.30 | 40.67 | - 0.92 |
| ESO 565−G019 | -13.05 | -12.20 | -10.71 | -11.13 | 43.70 | 43.76 | 43.74 | 43.26 | 40.85 | ⋯ |
| Mrk 3 | -12.23 | -11.21 | -10.06 | -10.22 | 43.15 | 43.37 | 43.53 | 43.33 | 41.32 | - 1.58 |
| NGC 1194 | -13.28 | -11.94 | -10.64 | -10.80 | 42.45 | 42.72 | 42.92 | 42.63 | 40.75 | - 2.49 |
| NGC 3393 | -12.59 | -12.58 | -10.95 | -10.98 | 44.92 | 44.67 | 44.29 | 44.32 | 40.41 | 0.70 |
| NGC 4945 | -12.18 | -11.54 | -09.81 | -10.03 | 43.57 | 43.63 | 43.62 | 43.45 | 39.41 | 0.53 |
| NGC 5728 | -12.86 | -11.84 | -10.40 | -10.42 | 42.82 | 43.06 | 43.22 | 43.29 | 40.64 | - 1.74 |
| NGC 6552 | -13.45 | -12.19 | -10.98 | -11.09 | 42.36 | 42.73 | 43.07 | 42.95 | 41.30 | ⋯ |
| NGC 7130 | -12.62 | -12.65 | -11.20 | -11.15 | 44.13 | 43.87 | 43.50 | 43.61 | 40.41 | - 0.42 |
| NGC 7582 | -12.37 | -11.59 | -10.42 | -10.40 | 41.92 | 42.11 | 42.21 | 42.44 | 40.13 | - 2.03 |

Note. (1) Galaxy name. (2)–(5) Logarithmic observed flux in 0.5–2.0 keV (*Suzaku*/BIXIS), 2–10 keV (*Suzaku*/FIXISs), 10–50 keV (*Suzaku*/HXD) and 10–50 keV (*Swift*/BAT) in units of erg cm$^{-2}$ s$^{-1}$. (6)–(9) Logarithmic absorption corrected luminosity in the same energy as (2)–(5) in units of erg s$^{-1}$. (10) Logarithmic Fe K$\alpha$ luminosity. (11) Logarithmic Eddington ratio. We define the Eddington luminosity as $L_{\text{Edd}} = 1.26 \times 10^{38} M_{\text{BH}}/M_\odot$ and adopt the 2–10 keV to bolometric correction factor of 20.

TABLE 9
Fluxes and Luminosities with the Ikeda2 model

| (1) Galaxy Name | (2) log $F^{Suzaku}_{0.5-2}$ | (3) log $F^{Suzaku}_{2-10}$ | (4) log $F^{Suzaku}_{10-50}$ | (5) log $F^{Swift}_{10-50}$ | (6) log $L^{Suzaku}_{0.5-2}$ | (7) log $L^{Suzaku}_{2-10}$ | (8) log $L^{Suzaku}_{10-50}$ | (9) log $L^{Swift}_{10-50}$ | (10) log $L_{FeK\alpha}$ | (11) log $\lambda_{Edd}$ |
|---|---|---|---|---|---|---|---|---|---|---|
| CGCG 420−015 | -12.69 | -11.75 | -10.89 | -10.79 | 44.01 | 43.86 | 43.60 | 43.86 | 41.34 | - 0.99 |
| ESO 137−G034 | -12.85 | -12.27 | -11.16 | -10.92 | 42.39 | 42.38 | 42.28 | 42.75 | 40.23 | - 1.97 |
| ESO 323−G032 | -13.28 | -12.40 | -11.05 | -11.17 | 43.66 | 43.52 | 43.29 | 43.19 | 40.70 | - 0.94 |
| ESO 565−G019 | -13.05 | -12.20 | -10.71 | -11.14 | 43.64 | 43.73 | 43.74 | 43.14 | 40.86 | ... |
| Mrk 3 | -12.40 | -11.21 | -10.06 | -10.24 | 43.23 | 43.55 | 43.80 | 43.67 | 41.34 | - 1.34 |
| NGC 1194 | -12.42 | -11.93 | -10.64 | -10.81 | 42.86 | 43.16 | 43.39 | 43.25 | 40.73 | - 1.90 |
| NGC 3393 | -12.60 | -12.54 | -10.94 | -10.97 | 43.46 | 43.45 | 43.34 | 43.41 | 40.41 | - 0.49 |
| NGC 4945 | -12.24 | -11.52 | -09.80 | -10.02 | 42.48 | 42.79 | 43.03 | 42.67 | 39.44 | - 0.51 |
| NGC 5728 | -12.88 | -11.84 | -10.40 | -10.42 | 42.68 | 42.94 | 43.13 | 43.23 | 40.66 | - 1.83 |
| NGC 6552 | -12.64 | -12.66 | -11.19 | -11.15 | 43.16 | 43.47 | 43.70 | 43.59 | 41.34 | ... |
| NGC 7130 | -12.64 | -12.66 | -11.19 | -11.15 | 43.39 | 43.13 | 42.76 | 43.03 | 40.40 | - 1.01 |
| NGC 7582 | -12.37 | -11.58 | -10.43 | -10.40 | 42.12 | 42.25 | 42.31 | 43.01 | 40.14 | - 1.41 |

Note. (1) Galaxy name. (2)–(5) Logarithmic observed flux in 0.5–2.0 keV (Suzaku/BIXIS), 2–10 keV (Suzaku/FIXISs), 10–50 keV (Suzaku/HXD) and 10–50 keV (Swift/BAT) in units of erg cm$^{-2}$ s$^{-1}$. (6)–(9) Logarithmic absorption corrected luminosity in the same energy as (2)–(5) in units of erg s$^{-1}$. (10) Logarithmic Fe K$\alpha$ luminosity. (11) Logarithmic Eddington ratio. We define the Eddington luminosity as $L_{Edd} = 1.26 \times 10^{38} M_{BH}/M_\odot$ and adopt the 2–10 keV to bolometric correction factor of 20.

TABLE 10
Spectral Curvature and MIR Luminosity

| (1) Galaxy Name | (2) Spectral Curvature | (3) log $\lambda L_{12\ \mu m}$ | (4) log $\lambda L_{12\ \mu m}$ Ref. |
| --- | --- | --- | --- |
| CGCG 420–015 | $0.37 \pm 0.24$ | 44.12 | (1) |
| ESO 137–G034 | $0.15 \pm 0.35$ | 42.73 | (1) |
| ESO 323–G032 | $1.12 \pm 0.78$ | 42.97 | (2) |
| ESO 565–G019 | $0.88 \pm 0.48$ | 43.13 | (1) |
| Mrk 3 | $0.54 \pm 0.05$ | 43.71 | (2) |
| NGC 1194 | $0.82 \pm 0.19$ | 43.45 | (2) |
| NGC 3393 | $1.08 \pm 0.32$ | 42.88 | (2) |
| NGC 4945 | $1.10 \pm 0.04$ | 39.95 | (2) |
| NGC 5728 | $0.36 \pm 0.12$ | 42.48 | (2) |
| NGC 6552 | $0.49 \pm 0.29$ | 43.78 | (1) |
| NGC 7130 | $0.39 \pm 0.48$ | 43.18 | (2) |
| NGC 7582 | $0.52 \pm 0.07$ | 42.85 | (2) |

Note. (1) Galaxy name. (2) Spectral curvature. (3) Logarithmic 12 $\mu$m luminosity. (4) Reference of the 12 $\mu$m luminosity.
References. (1) The data taken from the ALLWISE Source Catalog (Wright et al. 2010). (2) Asmus et al. (2015).

Since the quality of the X-ray spectrum of this object is very high, we add emission lines[8] reported by Awaki et al. (2008) to the spectral models. The central energy and width of these lines are fixed at the literature value and 10 eV, respectively, whereas the normalizations are set free. No apec model is considered.

We are unable to reproduce the spectra with the Ikeda1 model ($\chi^2/\mathrm{dof} = 1002.4/658.0$). In the Ikeda2 model, we fix the torus opening angle and inclination angles at 50 and 51 degrees (Ikeda et al. 2009), respectively. This yields a much better fit than the Ikeda1 model ($\chi^2/\mathrm{dof} = 766.7/659.0$). We obtain $N_\mathrm{H}^\mathrm{Dir} = 1.11^{+0.09}_{-0.04} \times 10^{24}$ cm$^{-2}$ and $\Gamma = 1.60^{+0.05}_{-0.02}$ with the model. Our results are consistent with the previous *Suzaku* study by Ikeda et al. (2009), who obtained $N_\mathrm{H}^\mathrm{Dir} = 1.1 \times 10^{24}$ cm$^{-2}$ and $\Gamma = 1.82$ with the Ikeda torus model. Using the same *Suzaku* data, Yaqoob et al. (2015) obtained $N_\mathrm{H}^\mathrm{Dir} = 0.90^{+0.01}_{-0.01} \times 10^{24}$ cm$^{-2}$ and $\Gamma = 1.47^{+0.01}_{-0.01}$ with the MYTorus model, and Guainazzi et al. (2016) obtained $N_\mathrm{H} = 0.86^{+0.01}_{-0.01} \times 10^{24}$ cm$^{-2}$ and $\Gamma = 1.76^{+0.02}_{-0.01}$ using *NuSTAR* data taken on 2014/09/07 (the closest epoch to the Suzaku observation among their 9 observation epochs) with the Ikeda torus model. These values are slightly different from our results. The discrepancy is most probably due to the different torus geometry ($\theta_\mathrm{open} = 66$ degree and $\theta_\mathrm{incl} = 70$ degree are assumed in Guainazzi et al. 2016), and/or possible time variability (Guainazzi et al. 2016).

[8] O$_\mathrm{VII}$ K$\alpha$ (0.56 keV), O$_\mathrm{VIII}$ Ly$\alpha$ (0.65 keV), O$_\mathrm{VII}$ RRC (0.74 keV), O$_\mathrm{VIII}$ Ly$\alpha$ (0.87 keV), Ne$_\mathrm{IX}$ K$\alpha$ (0.92 keV), Ne$_\mathrm{X}$ Ly$\alpha$ (1.02 keV), Fe$_\mathrm{XXII}$ Ly$\alpha$ (1.05 keV), Fe$_\mathrm{XXIII}$ Ly$\alpha$ (1.17 keV), Mg K$\alpha$ (1.25 keV), Mg$_\mathrm{IX}$ K$\alpha$ (1.33 keV) Mg$_\mathrm{XII}$ Ly$\alpha$ (1.47 keV), Si$_\mathrm{XIV}$ Ly$\alpha$ (2.01 keV), S K$\alpha$ (2.31 keV), S$_\mathrm{XV}$ Ly$\alpha$ (2.45 keV), S$_\mathrm{XVI}$ Ly$\alpha$ (2.62 keV), Fe K$\alpha$ (6.40 keV), Fe$_\mathrm{XXV}$ Ly$\alpha$ (6.64 keV), Fe K$\beta$ (7.06 keV) and Ni K$\alpha$ (7.48 keV)

### 4.6. *NGC 1194*

The *Suzaku* spectra are reported for the first time here. We detect Fe K$\alpha$ ($EW = 0.57^{+0.55}_{-0.24}$ keV) and K$\beta$ ($EW = 0.16^{+0.08}_{-0.08}$ keV) emission lines. No apec component is required. The baseline models are able to reproduce the spectra, while the Ikeda torus models give worse fits. Nevertheless, we obtained $N_\mathrm{H}^\mathrm{Dir} = 1.55^{+1.48}_{-0.41} \times 10^{24}$ cm$^{-2}$ and $\Gamma = 1.63^{+0.25}_{-0.12}$ with the Ikeda2 model. They are consistent with those with the baseline models.

### 4.7. *NGC 3393*

All the models with two apec components provide acceptable fits. We detect a Fe K$\alpha$ ($EW = 2.13^{+33.9}_{-0.92}$ keV) emission line. We obtain $N_\mathrm{H}^\mathrm{Dir} = 2.78(> 1.48) \times 10^{24}$ cm$^{-2}$ and $\Gamma = 2.11(> 1.72)$ with the Ikeda2 model. Our results are consistent with the results using the *NuSTAR* data with the MYTorus model (Koss et al. 2015), $N_\mathrm{H}^\mathrm{Dir} = 1.85 \times 10^{24}$ cm$^{-2}$ and $\Gamma = 1.7^{+0.2}_{-0.2}$.

### 4.8. *NGC 4945*

NGC 4945 is a nearby Seyfert 2 galaxy and one of the brightest AGNs above 10 keV. This object has been observed by many X-ray satellites since *Ginga*, such as *RXTE* (Madejski et al. 2000), *BeppoSAX* (Guainazzi et al. 2000), *XMM-Newton* (Schurch et al. 2002), *Chandra* (Done et al. 2003) *Swift* (Tueller et al. 2008, 2010; Winter et al. 2008, 2009a), *Suzaku* (Itoh et al. 2008; Yaqoob 2012) and *NuSTAR* (Puccetti et al. 2014; Brightman et al. 2016). The soft X-ray spectrum of NGC 4945 is known to be very complex, originating from the starburst activities in the host galaxy (Itoh et al. 2008; Yaqoob 2012). Thus, we exclude the 0.5–2.0 keV from our spectral analysis to focus on the AGN component. Following Itoh et al. (2008), we add emission lines of Fe K$\alpha$ (6.40 keV), Fe$_\mathrm{XXV}$ Ly$\alpha$ (6.64 keV), Fe K$\beta$ (7.06 keV) and Ni Kα (7.48 keV) in the spectral models. All models with these emission lines (with no apec component) reproduce the spectra reasonably well. The best-fit Ikeda1 model predicts that most of the hard X-ray emission comes from the torus reflection component. This picture seems incompatible with the observed fast time variability in the hard X-ray band (Itoh et al. 2008; Yaqoob 2012). In fact, to explain the time variability, Yaqoob (2012) concluded that the hard X-ray emission was dominated by the direct component, indicating a clumpy torus where a high column-density cloud is present in the line-of-sight. Hence, we consider that the fitting results of the Ikeda1 model for this target is unphysical and we discuss the results obtained by the Ikeda2 model.

We obtain $N_{\rm H}^{\rm Dir} = 3.45^{+0.38}_{-0.32} \times 10^{24}$ cm$^{-2}$ and $\Gamma = 1.62^{+0.04}_{-0.04}$ with the Ikeda2 model. Using the same *Suzaku* data, Itoh et al. (2008) obtained $N_{\rm H} = 5.3^{+0.4}_{-0.9} \times 10^{24}$ cm$^{-2}$ and $\Gamma = 1.6^{+0.1}_{-0.2}$, and Yaqoob (2012) obtained $N_{\rm H} = 4.00^{+0.10}_{-0.07} \times 10^{24}$ cm$^{-2}$ with the MYTorus model where the line-of-sight absorption is decoupled from the torus parameters. Note that they utilized the *Suzaku*/HXD-GSO data instead of the *Swift*/BAT spectrum, which may cause the small difference in $N_{\rm H}$. Using the *NuSTAR* data, Puccetti et al. (2014) obtained $N_{\rm H}^{\rm Dir} = 3.5^{+0.2}_{-0.2} \times 10^{24}$ cm$^{-2}$ and $\Gamma = 1.77^{+0.09}_{-0.09}$ in the low state with the MYTorus model. These results are consistent with our findings.

### 4.9. *NGC 5728*

All the models with one apec component provide acceptable fits. We detect Fe Kα ($EW = 0.87^{+0.80}_{-0.26}$ keV) and Kβ ($EW = 0.16^{+0.08}_{-0.08}$ keV) emission lines. With the Ikeda2 model, we obtain $N_{\rm H}^{\rm Dir} = 1.69^{+1.45}_{-0.53} \times 10^{24}$ cm$^{-2}$ and $\Gamma = 1.69^{+0.14}_{-0.14}$. Our results are consistent with the *Suzaku* results (Comastri et al. 2010). They obtained $N_{\rm H} = 1.40^{+0.05}_{-0.08} \times 10^{24}$ cm$^{-2}$ and $\Gamma = 1.68^{+0.05}_{-0.04}$ with the reflection-dominant model.

### 4.10. *NGC 6552*

The *Suzaku* broadband X-ray spectra are reported for the first time here. We detect Fe Kα ($EW = 1.24^{+2.37}_{-0.37}$ keV) and Kβ ($EW = 0.16^{+0.09}_{-0.10}$ keV) lines. The baseline models with one apec component reproduces the spectra reasonably well. The Ikeda torus models (with one apec component) give worse fits. With the Ikeda2 model, we obtain $N_{\rm H}^{\rm Dir} = 1.74^{+1.56}_{-0.99} \times 10^{24}$ cm$^{-2}$ and $\Gamma = 1.62(< 1.76)$.

### 4.11. *NGC 7130*

The *Suzaku* broadband X-ray spectra are reported for the first time here. All the models with one apec component provide acceptable fits. An Fe Kα ($EW = 1.34^{+8.07}_{-1.22}$ keV) line is detected. The best fit parameters with the Ikeda2 model are $N_{\rm H}^{\rm Dir} = 1.69^{+2.82}_{-0.54} \times 10^{24}$ cm$^{-2}$ and $\Gamma = 2.50(> 2.27)$. Our results are consistent with the *Chandra* results by González-Martín et al. (2009), who obtained $N_{\rm H} = 0.86^{+0.74}_{-0.26} \times 10^{24}$ cm$^{-2}$ and $\Gamma = 2.66^{+0.16}_{-0.12}$.

### 4.12. *NGC 7582*

NGC 7582 is known to show variable absorption (Bianchi et al. 2009). The *Suzaku* spectra analyzed here correspond to the epoch with the deepest obscuration (S4) among the four datasets reported by Bianchi et al. (2009). All the models with one apec components give reasonable description of the spectra. We detect an Fe Kα ($EW = 0.44^{+0.15}_{-0.09}$ keV) line. With the Ikeda2 model, we obtain $N_{\rm H}^{\rm Dir} = 3.23^{+1.33}_{-1.78} \times 10^{24}$ cm$^{-2}$ and $\Gamma = 1.88^{+0.11}_{-0.12}$. Piconcelli et al. (2007) obtained $N_{\rm H} = 1.29^{+0.06}_{-0.07} \times 10^{24}$ cm$^{-2}$ and $\Gamma = 1.93^{+0.01}_{-0.01}$ using the *XMM-Newton* data with an analytical model, Bianchi et al. (2009) obtained $N_{\rm H} = 1.20^{+0.20}_{-0.20} \times 10^{24}$ cm$^{-2}$ and $\Gamma = 1.92^{+0.24}_{-0.16}$ using the *Suzaku* data with an analytical model, and (Rivers et al. 2015) obtained $N_{\rm H}^{\rm Dir} = 1.04^{+0.13}_{-0.13} \times 10^{24}$ cm$^{-2}$ and $\Gamma = 1.82^{+0.07}_{-0.07}$ using the *NuSTAR* data with MYTorus model. Our $N_{\rm H}$ value is larger than that found by previous studies, because our best-fit Ikeda2 model shows a reflection-dominant spectrum (Figure 3).

## 5. DISCUSSION

We have presented the broadband X-ray spectra of 12 CTAGNs observed with *Suzaku* and *Swift*. They are one of the best-quality datasets covering the 0.5–100.0 keV range ever observed from local CTAGNs. We have shown that the spectra can be successfully described with analytic models with the pexrav code (Baseline1 and Baseline2) and numerical models with the Ikeda et al. (2009) torus model (Ikeda1 and Ikeda2) except for a few cases. The four models give consistent results on the line-of-sight hydrogen column density and photon index, confirming the heavy obscuration of all the targets. We find, however, that the estimates of the intrinsic luminosity and the decomposition of the transmitted and reflection components are not always unique, depending on the assumed spectral model (i.e., the torus geometry). We discuss the details in the following.

### 5.1. *Comparison of Intrinsic Luminosities among Models*

Estimating true intrinsic X-ray luminosities of obscured AGNs is an important yet nontrivial issue. It depends on the decomposition of multiple components, particularly the torus reflection, as previously discussed (see e.g., Ikeda et al. 2009; Murphy & Yaqoob 2009). Murphy & Yaqoob (2009) pointed out that the effect of Compton scattering along the line-of-sight is important to derive correct intrinsic luminosities in CTAGNs. They also showed that the spectral shape of the reflection continuum in the MYTorus model is quite different from that calculated with the pexrav model both at energies below and above 10 keV (see their Figure 7), which could affect the spectral fitting results. In this subsection, on the basis of our uniform spectral analysis of a large sample of local CTAGNs, we discuss how estimates of the intrinsic luminosities would be changed by spectral models.

We compare the intrinsic 10–50 keV luminosities obtained with different models. Here we adopt the 10–50 keV instead of the canonical 2–10 keV because the transmitted component is directly measured only at energies above 10 keV in CTAGNs. Figures 5 (a), (b), (c) and (d) compare the luminosities obtained from the (a) Baseline1 and Baseline2 models, (b) Ikeda1 and Ikeda2 models (except for Mrk 3 and NGC 4945, whose Ikeda1 fit is poor or unphysical), (c) Baseline1 and Ikeda2 models and (d) Baseline2 and Ikeda2 models, respectively. Figures 5 (e) and (f) plot the logarithmic luminosity differences (e) between the Baseline1 and Ikeda2 models and (f) between the Baseline2 and Ikeda2 models, respectively, as a function of line-of-sight hydrogen column density.

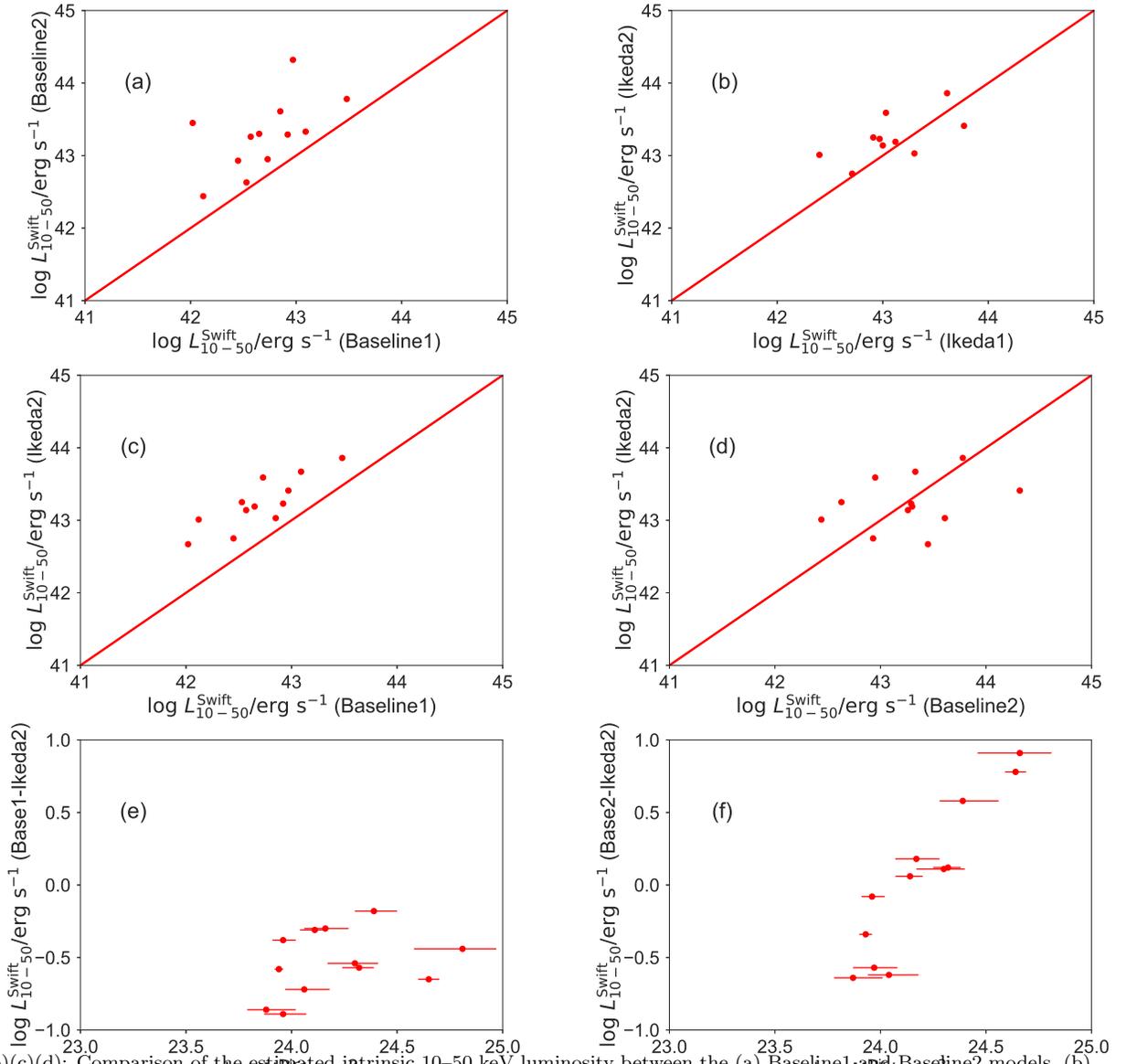

FIG. 5.— (a)(b)(c)(d): Comparison of the estimated intrinsic 10–50 keV luminosity between the (a) Baseline1 and Baseline2 models, (b) Ikeda1 and Ikeda2 models, (c) Baseline1 and Ikeda2 models and (d) Baseline2 and Ikeda2 models. The two compared luminosities are equal at the red line. (e)(f): the logarithmic luminosity difference between the (e) Baseline1 and Ikeda2 models plotted against the column density of the Baseline1 model and (f) Baseline2 and Ikeda2 models plotted against the column density of the Baseline2 model.

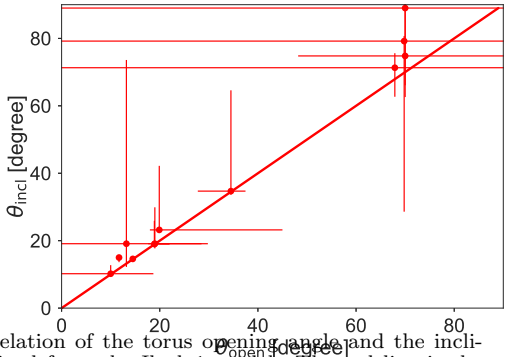

FIG. 6.— Correlation of the torus opening angle and the inclination angle derived from the Ikeda1 model. The red line is the equal line.

The Baseline2 model always gives higher luminosities than the Baseline1 model by $\sim 0.5$ dex (Figure 5 (a)). This is due to the Compton scattering term. The Baseline2 model assumes an extreme case where the absorber is located only along the line of sight, whereas the Baseline1 model corresponds to the other extreme case where photons scattering into the line-of-sight from other directions completely cancel out the the scattered-out photons. In Figure 5 (b), the luminosity differences be-tween the Ikeda1 and Ikeda2 models are less noticeable, although the Ikeda2 model tends to give higher luminosities than the Ikeda1 model by ∼0.2 dex. This is because the Ikeda1 model often favors a solution of a small opening angle (Section 5.2), which predicts strong reflection components and hence a weaker transmitted component. We consider that the torus geometry derived from the Ikeda1 model may be artificial in several cases (Section 5.2).

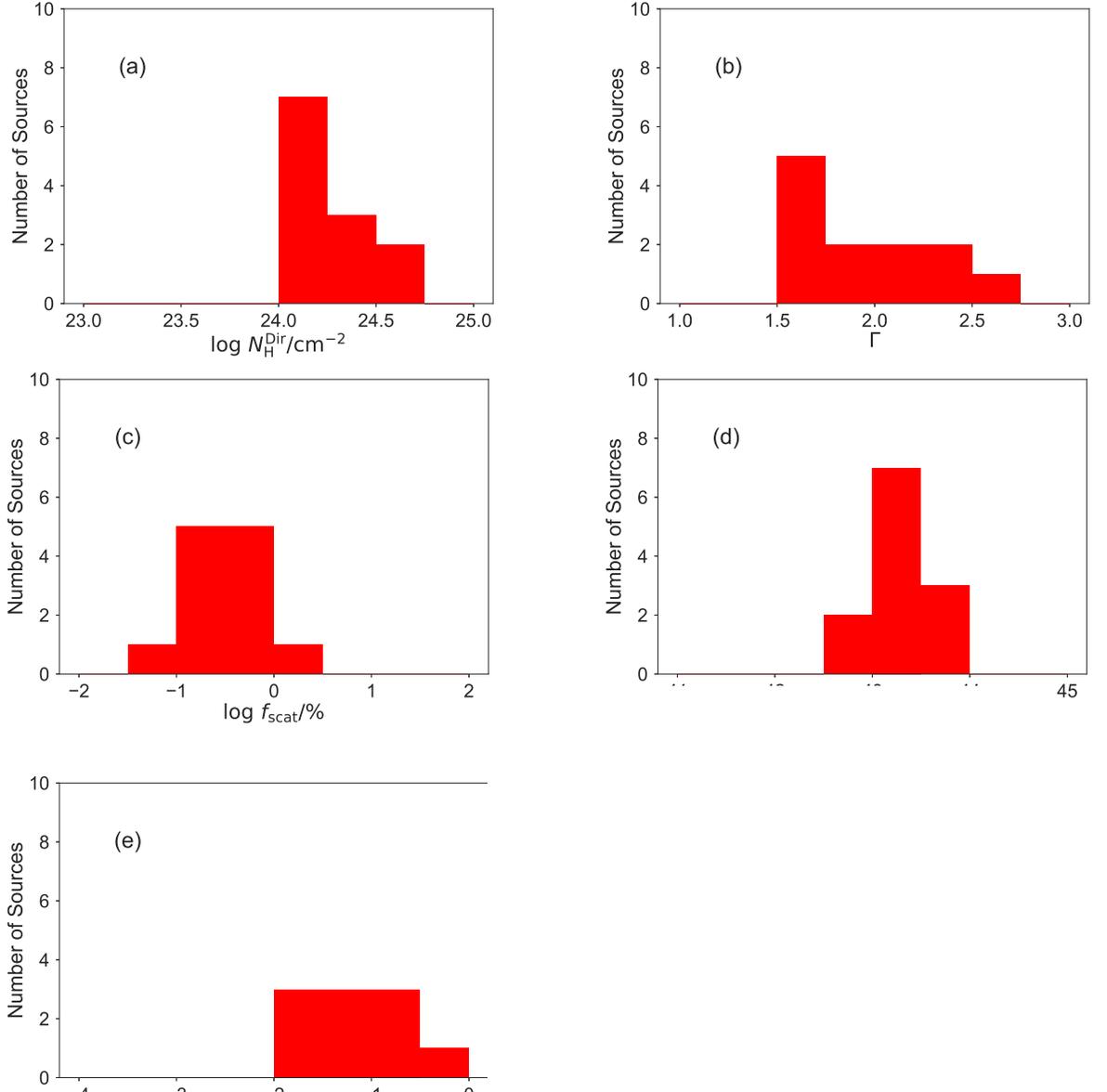

FIG. 7.— Histograms of the (a) line-of-sight hydrogen column density, (b) photon index, (c) scattered fraction, (d) intrinsic 10–50 keV luminosity and (e) Eddington ratio. They are all based on the Ikeda2 model.

We find that the Baseline1 model always underestimates the true 10–50 keV luminosities (Figure 5 (c) and (e)). This result is consistent with the argument by Murphy & Yaqoob (2009). In fact, Masini et al. (2017) found the same trend by applying MYTorus model to the Phoenix galaxy. Similarly, Ikeda et al. (2009) found that the intrinsic luminosity of Mrk 3 obtained with the Ikeda torus model is about 1.3 times larger that with an analytical model without the line-of-sight Compton scattering. On the other hand, the Baseline2 model overestimates them at high column densities (Figure 5 (d) and (f)). The real physical situation would be between the two cases. Hereafter, we refer to the results of the Ikeda2 model as the most realistic case.

### 5.2. Torus Parameters

Figure 6 plots the correlation between the half-opening angle and inclination obtained from the Ikeda1 model fit. Remarkably, in most cases the differences between the two angles are very small, even < 1 degree. This happens because the spectra require a large amount of unabsorbed reflection component that is noticeable below the Fe K edge (7.1 keV). If we took these results naively, it would mean that the line-of-sight is intercepted by the boundary edge of the torus. However, considering the fact that similar results have been obtained from a large fraction of heavily obscured AGNs (e.g., NGC 2273, Awaki et al. 2009; NGC 3081, Eguchi et al. 2011; 3C 403, Tazaki et al. 2011) such a picture would be unrealistic. Our results reinforce the interpretation by Tanimoto et al. (2016) that these spectral features are caused by clumpy tori (Krolik & Begelman 1988; Nenkova et al. 2008a,b; Kawaguchi & Mori 2011), from which unabsorbed reflection components are easily observable (Liu & Li 2014; Furui et al. 2016).

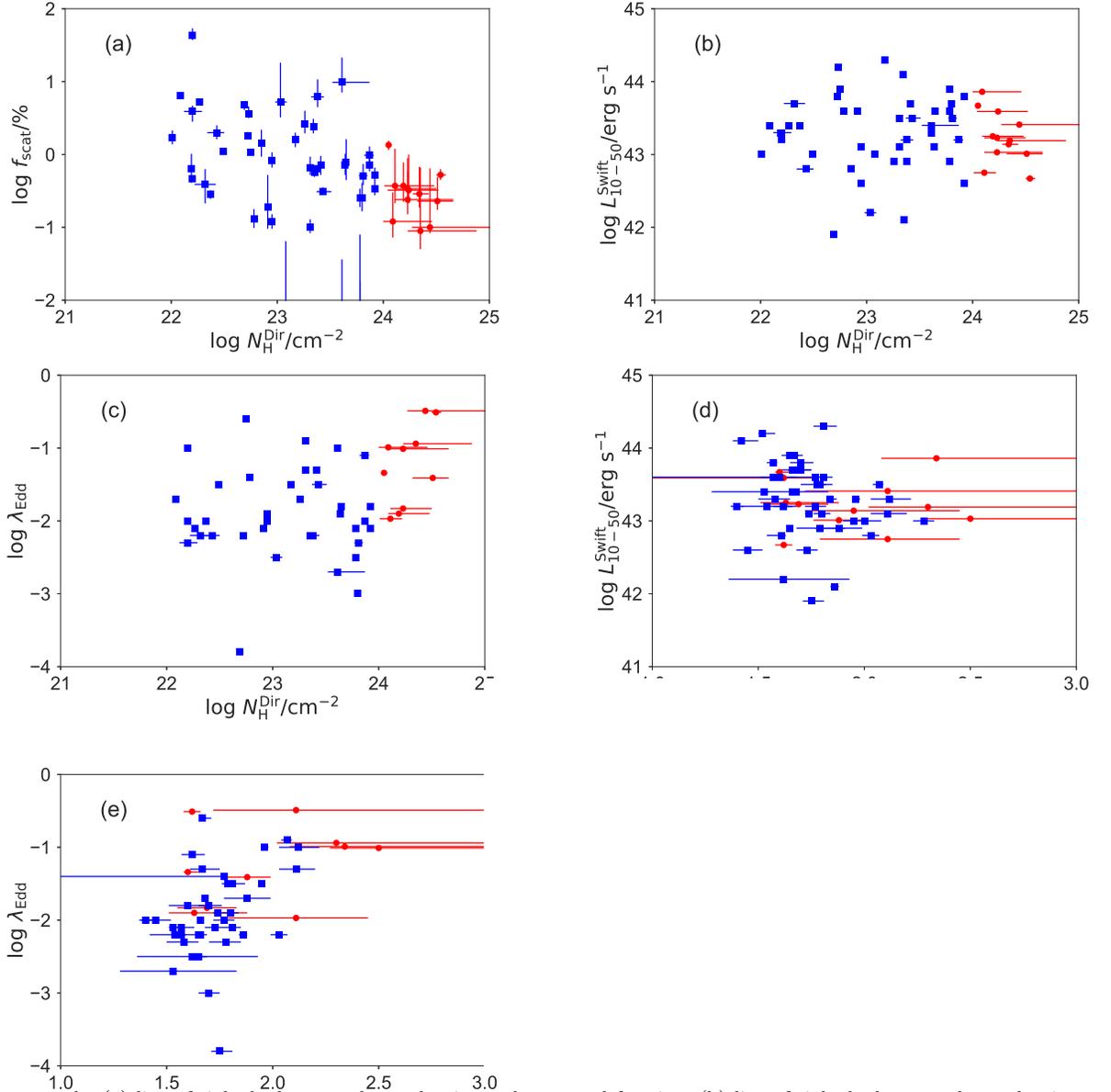

FIG. 8.— Plot between the (a) line-of-sight hydrogen column density and scattered fraction, (b) line-of-sight hydrogen column density and 10–50 keV luminosity, (c) line-of-sight hydrogen column density and Eddington ratio, (d) photon index and 10–50 keV luminosity and (e) photon index and Eddington ratio. The blue squares and red circles correspond to the Compton-thin AGNs of Kawamuro et al. (2016a) and our sample, respectively.

We thus consider that the fitting results with the Ikeda1 model are artificial in several cases and should not be taken at their face values. For instance, we obtain small inclination angles from three water megamaser AGNs in our sample, NGC 1194, NGC 3393 and NGC 4945, which must be observed close to edge-on in reality. In the Ikeda2 model, we have assumed an inclination of 70 degrees (a practical upper limit in the Ikeda model) except for Mrk 3, which would be more proper for CTAGNs. We then obtain best-fit half-opening angles of $\approx$ 65–70 degrees in order to reproduce the unabsorbed reflection component. The column density at the equatorial plane ($N_{\rm H}^{\rm Equ}$) is determined to account for the intensities of the reflection continuum. Even with an apparently large opening angle (e.g., > 65 degrees), we do not rule out the possibility that Compton-thin material ($N_{\rm H} \leq 10^{24}$ cm$^{-2}$) is present in the torus-hole region (i.e., at inclinations lower than this half-opening angle) and "buried" the nucleus, because it would not affect significantly the broadband spectra except for the scattered component. In fact, such a geometry is proposed for the CTAGN in UGC 5101 (Oda et al. 2017) and possibly for NGC 4945.

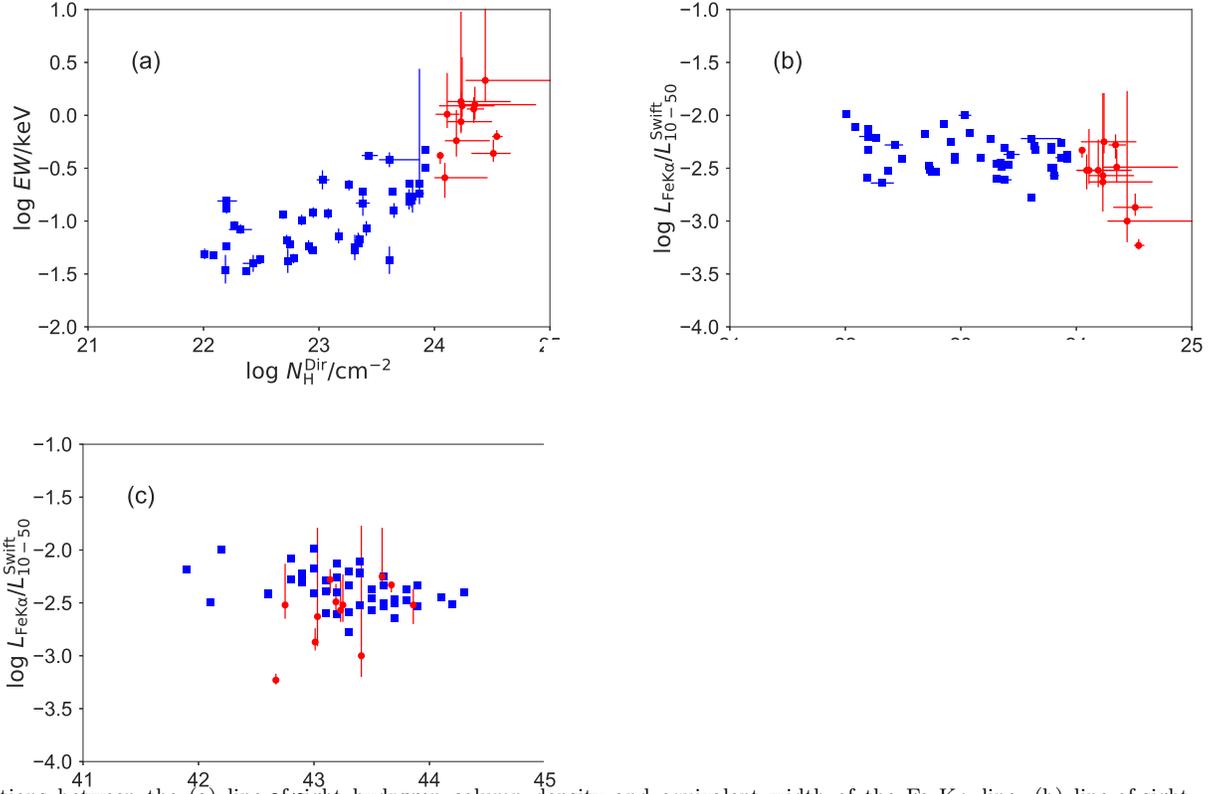

FIG. 9.— Relations between the (a) line-of-sight hydrogen column density and equivalent width of the Fe Kα line, (b) line-of-sight hydrogen column density and Fe Kα to 10–50 keV luminosity ratio, (c) 10–50 keV luminosity and Fe Kα to 10–50 keV luminosity ratio. The blue squares and red circle correspond to the Compton-thin AGNs of Kawamuro et al. (2016a) and our sample, respectively.

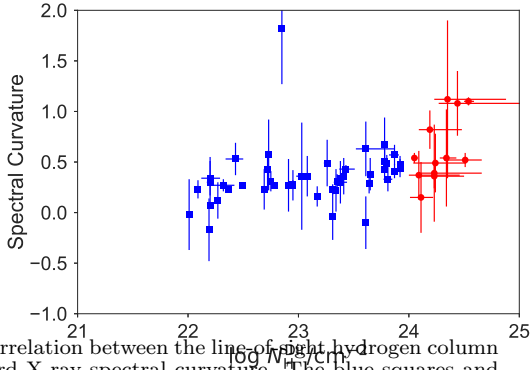

FIG. 10.— Correlation between the line-of-sight hydrogen column density and hard X-ray spectral curvature. The blue squares and red circle correspond to the Compton-thin AGNs of Kawamuro et al. (2016a) and our sample, respectively.

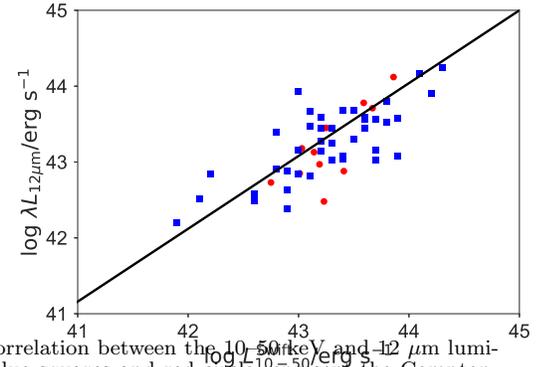

FIG. 11.— Correlation between the 10–50 keV and 12 μm luminosities. The blue squares and red circle represent the Compton-thin AGNs of Kawamuro et al. (2016a) and our sample, respectively. The black line shows the result of Ichikawa et al. (2017) obtained from the whole unbeamed AGN sample in the *Swift*/BAT 70-month catalog.

### 5.3. *Correlations among Basic Parameters*

Figure 7 shows the distribution of the (a) line-of-sight hydrogen column density, (b) photon index, (c) scattered fraction, (d) 10–50 keV luminosity and (e) Eddington ratio, all based on the Ikeda2 model fit. Figure 7(a) confirms that all the objects are CTAGNs. We find that 10 AGNs show a scattered fraction smaller than 0.5% (Figure 7(c)), which are referred to as new type AGNs by Ueda et al. (2007).

Figure 8 shows the relation between the (a) hydrogen column density and scattered fraction, (b) hydrogen column density and 10–50 keV luminosity, (c) hydrogen column density and Eddington ratio, (d) photon index and 10–50 keV luminosity and (e) photon index and Eddington ratio. For comparison, hereafter we also plot the data of Compton-thin AGNs obtained by Kawamuro et al. (2016a) with blue squares; we refer to their results of the *intrinsic* (de-absorbed) 10–50 keV luminosity that does not include the reflection component, instead of the *observed* 10–50 keV luminosity, for consistency with our plot.

Figure 8 (a) suggests that the fraction of small scattered-fraction AGNs is larger in more heavily obscured AGNs, particularly in CTAGNs, even though the samples are not statistically complete. This implies that a majority of CTAGNs are deeply buried in geometrically thick tori, that is, there is a link between the column density and geometry (covering fraction) in AGN tori (Kawamuro et al. 2016a). Similar implication is also obtained from a sample of distant AGNs (Brightman & Ueda 2012). As noted above, this interpretation does not conflict with the relatively large opening angles derived from the Ikeda2 model fit if the torus-hole region is covered by Compton-thin material that only affects the scattered component below several keV.

For this sample of obscured AGN we do not find any statistically significant correlation between the column density and the X-ray luminosity (Figure 8(b)) or between the column density and the Eddington ratio (Figure 8(c)). Our CTAGN sample is located within the scatter of the $\Gamma$ versus X-ray luminosity relation (Figure 8(d)) and the $\Gamma$ versus Eddington ratio plot (Figure 8(e)) obtained from the Compton-thin AGNs. We note that NGC 4945 may be an outlier showing a somewhat flat photon index ($\Gamma \simeq 1.6$) at a high Eddington ratio ($\log \lambda_{\mathrm{Edd}} = -0.10$).

### 5.4. Fe Kα Line

We detected a strong Fe Kα emission line (EW = 0.26–2.13 keV) from all targets. Figure 9 plots the (a) equivalent width of Fe Kα line against the column density, (b) luminosity ratio between the Fe Kα line and intrinsic 10–50 keV continuum against the column density and (c) that against the 10–50 keV luminosity.

Figure 9(a) confirms the general correlation between the line-of-sight hydrogen column density and the equivalent width of the Fe Kα line as previously reported (e.g., Turner et al. 1998; Turner & Miller 2009; Fukazawa et al. 2011). The correlation at $\log N_\mathrm{H} > 23.0$ cm$^{-2}$ is mainly caused by the attenuation of the transmitted component by photoelectric absorption at the same energy, which makes the equivalent width of the Fe Kα line larger. This indicates that the Fe Kα equivalent width is a good indicator to identify CTAGNs.

The ratio between the Fe Kα and hard X-ray (> 10 keV) luminosities is proposed to be a better indicator than the equivalent width to constrain the solid angle of the torus without being affected by the continuum absorption (Ricci et al. 2014). As noticed from Figure 9(b), this luminosity ratio tends to decrease toward larger hydrogen column densities. This effect is attributable to self-absorption of the emitted Fe Kα line by the near-side torus (Ricci et al. 2014). It would explain the reason why our CTAGN sample, though limited in number, apparently does not follow the anti-correlation between the $L_{\mathrm{FeK}\alpha}/L_{10-50}$ ratio and the 10–50 keV luminosity found for Compton-thin AGNs by Ricci et al. (2014) and Kawamuro et al. (2016a) (Figure 9(c)).

### 5.5. *Spectral Curvature*

Koss et al. (2016) propose a new method that uses the "spectral curvature" above 10 keV to identify CTAGNs with *Swift*/BAT data. They focus on data below 50 keV because this energy band shows the strongest difference in the curvature compared with unobscured AGNs. The definition of the spectral curvature is

$$\mathrm{SC_{BAT}} = \frac{-3.42A - 0.82B + 1.65C + 3.58D}{\text{Total Rate}} \quad (6)$$

where A, B, C and D refer to the 14–20 keV, 20–24 keV, 24–35 keV and 35–50 keV *Swift*/BAT count rates, respectively, and the total rate refers to the 14–50 keV count rate. The spectral curvature is calibrated so that a heavily CTAGN in an edge-on torus has a value of 1 ($N_\mathrm{H} = 5.0 \times 10^{24}$ cm$^{-2}$) and an unobscured AGN has a value of 0. Table 10 lists the spectral curvatures calculated from the *Swift*/BAT 70 month spectra for our CTAGN sample. Figure 10 plots the spectral curvature against the column density for our sample and for the Compton-thin AGN sample of Kawamuro et al. (2016a). We confirm a systematic trend that the spectral curvature increases with column density at $N_\mathrm{H} \geq 1.5 \times 10^{24}$ cm$^{-2}$, although the scatter is large. There is one outlier Compton-thin AGN (NGC 3431) that shows a large spectral curvature of ≈1.8. The reason is unclear but it may be simply due to statistical fluctuation if we consider the large error.

### 5.6. *Correlation of X-ray and MIR Luminosities*

We discuss the correlation between X-ray and mid-infrared (MIR) luminosities. The MIR emission of an AGN mainly comes from the dust in the torus heated by the primary radiation from the central engine. It is known that there is a good correlation between the X-ray and MIR luminosities in AGNs (Gandhi et al. 2009; Ichikawa et al. 2012; Asmus et al. 2015; Kawamuro et al. 2016a; Ichikawa et al. 2017). Table 10 summarizes the $\lambda L_{12\mu\mathrm{m}}$ luminosities of our sample obtained with the *WISE* observatory or ground-based facilities (Wright et al. 2010; Asmus et al. 2015). Figure 11 shows correlation between the 10–50 keV luminosity and the 12 $\mu$m luminosity for our CTAGNs and the Compton-thin AGNs in Kawamuro et al. (2016a), In this figure we also show the regression line derived by Ichikawa et al. (2017) by using a large (∼ 700) sample of *Swift*/BAT AGNs matched to the *WISE* catalog; here we convert the 14–195 keV luminosities into the 10–50 keV ones by assuming a photon index of 1.8.

We confirm that our sample generally follows the same correlation as for less obscured AGNs (Gandhi et al. 2009; Ichikawa et al. 2012; Asmus et al. 2015; Kawamuro et al. 2016a; Ichikawa et al. 2017). It is naively expected that deeply buried AGNs with geometrically thick tori may emit stronger MIR emission at the same hard X-ray luminosity. However, according to detailed theoretical calculations of the infrared radiation from clumpy tori (Stalevski et al. 2016), the MIR emission decreases with the inclination (see their Figures 8 and 11 for the face-on and edge-on cases). Thus, if many of our CTAGNs are observed close to edge-on, the two effects would canceled out each other. More detailed comparison between the infrared and X-ray properties of CTAGNs will be useful to reveal the geometry of their tori.

### 6. CONCLUSION

1. The estimate of the intrinsic luminosity of a Compton-thick AGN (CTAGN) strongly depends

on the spectral model. Applying Compton scattering to the transmitted component in an analytic model may largely overestimate the true luminosity at large column densities. The usage of Monte Carlo based model assuming a realistic geometry is required to estimate the intrinsic luminosity most reliably.

2. Uniform-density torus models tend to give a geometrical solution where the line-of-sight is intercepted near the edge of the torus, in order to explain a large amount of unabsorbed reflection components from the far-side torus. We interpret this as evidence of clumpiness in the torus.

3. A large fraction of the objects of our sample (10 out of 12) shows small scattering fractions ($< 0.5\%$). This implies that a majority of CTAGNs is deeply buried in geometrically thick tori, which might imply that there is a link between the column density and covering fraction in AGN tori.

4. We confirm the Fe K$\alpha$ equivalent width is a good indicator to identify CTAGNs without detailed spectral modeling.

5. The overall results confirm that the properties of hard X-ray selected CTAGNs can be understood as a smooth extension from Compton-thin AGNs with heavier obscuration; we find no evidence that they are distinct populations from less obscured AGNs.

We thank Poshak Gandhi for providing the HXD-PIN background spectrum of ESO 565–G019, and Franz Bauer for useful discussion. Part of this work was financially supported by the Grant-in-Aid for Scientific Research 17K05384 (Y.U.), 16K05296 (Y.T.) and 15H02070 (H.A. and Y.T.) and for JSPS Fellows for young researchers (A.T. and T.K.). We acknowledge financial support from the China-CONICYT fellowship (C.R.), FONDECYT 1141218 (C.R.) and Basal-CATA PFB-06/2007 (C.R.). This research has made use of the NASA/IPAC Infrared Science Archive, which is operated by the Jet Propulsion Laboratory, California Institute of Technology, under contract with the National Aeronautics and Space Administration.

## REFERENCES

Aird, J., Coil, A. L., Georgakakis, A., et al. 2015, MNRAS, 451, 1892
Ajello, M., Alexander, D. M., Greiner, J., et al. 2012, ApJ, 749, 21
Ajello, M., Rau, A., Greiner, J., et al. 2008, ApJ, 673, 96
Akylas, A., Georgakakis, A., Georgantopoulos, I., Brightman, M., & Nandra, K. 2012, A&A, 546, A98
Akylas, A., Georgantopoulos, I., Ranalli, P., et al. 2016, A&A, 594, A73
Anders, E., & Grevesse, N. 1989, Geochim. Cosmochim. Acta, 53, 197
Asmus, D., Gandhi, P., Hönig, S. F., Smette, A., & Duschl, W. J. 2015, MNRAS, 454, 766
Awaki, H., Terashima, Y., Higaki, Y., & Fukazawa, Y. 2009, PASJ, 61, S317
Awaki, H., Anabuki, N., Fukazawa, Y., et al. 2008, PASJ, 60, S293
Bauer, F. E., Arévalo, P., Walton, D. J., et al. 2015, ApJ, 812, 116
Baumgartner, W. H., Tueller, J., Markwardt, C. B., et al. 2013, ApJS, 207, 19
Beckmann, V., Gehrels, N., Shrader, C. R., & Soldi, S. 2006, ApJ, 638, 642
Beckmann, V., Soldi, S., Ricci, C., et al. 2009, A&A, 505, 417
Bianchi, S., Miniutti, G., Fabian, A. C., & Iwasawa, K. 2005, MNRAS, 360, 380
Bianchi, S., Piconcelli, E., Chiaberge, M., et al. 2009, ApJ, 695, 781
Brightman, M., & Nandra, K. 2011a, MNRAS, 413, 1206
—. 2011b, MNRAS, 414, 3084
Brightman, M., & Ueda, Y. 2012, MNRAS, 423, 702
Brightman, M., Masini, A., Ballantyne, D. R., et al. 2016, ApJ, 826, 93
Burlon, D., Ajello, M., Greiner, J., et al. 2011, ApJ, 728, 58
Comastri, A., Iwasawa, K., Gilli, R., et al. 2010, ApJ, 717, 787
Done, C., Madejski, G. M., Życki, P. T., & Greenhill, L. J. 2003, ApJ, 588, 763
Eguchi, S., Ueda, Y., Awaki, H., et al. 2011, ApJ, 729, 31
Eguchi, S., Ueda, Y., Terashima, Y., Mushotzky, R., & Tueller, J. 2009, ApJ, 696, 1657
Fukazawa, Y., Mizuno, T., Watanabe, S., et al. 2009, PASJ, 61, S17
Fukazawa, Y., Hiragi, K., Mizuno, M., et al. 2011, ApJ, 727, 19
Furui, S., Fukazawa, Y., Odaka, H., et al. 2016, ApJ, 818, 164
Gandhi, P., Hönig, S. F., & Kishimoto, M. 2015, ApJ, 812, 113
Gandhi, P., Horst, H., Smette, A., et al. 2009, A&A, 502, 457
Gandhi, P., Terashima, Y., Yamada, S., et al. 2013, ApJ, 773, 51
Gilli, R., Comastri, A., & Hasinger, G. 2007, A&A, 463, 79
González-Martín, O., Masegosa, J., Márquez, I., Guainazzi, M., & Jiménez-Bailón, E. 2009, A&A, 506, 1107
Guainazzi, M., Matt, G., Brandt, W. N., et al. 2000, A&A, 356, 463
Guainazzi, M., Risaliti, G., Awaki, H., et al. 2016, MNRAS, 460, 1954
Hopkins, P. F., Hernquist, L., Cox, T. J., et al. 2006, ApJS, 163, 1
Ichikawa, K., Ricci, C., Ueda, Y., et al. 2017, ApJ, 835, 74
Ichikawa, K., Ueda, Y., Terashima, Y., et al. 2012, ApJ, 754, 45
Ikeda, S., Awaki, H., & Terashima, Y. 2009, ApJ, 692, 608
Ishisaki, Y., Maeda, Y., Fujimoto, R., et al. 2007, PASJ, 59, 113
Itoh, T., Done, C., Makishima, K., et al. 2008, PASJ, 60, S251
Izumi, T., Kawakatu, N., & Kohno, K. 2016, ApJ, 827, 81
Kalberla, P. M. W., Burton, W. B., Hartmann, D., et al. 2005, A&A, 440, 775
Kawaguchi, T., & Mori, M. 2011, ApJ, 737, 105
Kawamuro, T., Ueda, Y., Tazaki, F., Ricci, C., & Terashima, Y. 2016a, ApJS, 225, 14
Kawamuro, T., Ueda, Y., Tazaki, F., & Terashima, Y. 2013, ApJ, 770, 157
Kawamuro, T., Ueda, Y., Tazaki, F., Terashima, Y., & Mushotzky, R. 2016b, ApJ, 831, 37
Khorunzhev, G. A., Sazonov, S. Y., Burenin, R. A., & Tkachenko, A. Y. 2012, Astronomy Letters, 38, 475
Konami, S., Matsushita, K., Gandhi, P., & Tamagawa, T. 2012, PASJ, 64, 117
Kormendy, J., & Ho, L. C. 2013, ARA&A, 51, 511
Koss, M., Trakhtenbrot, B., Ricci, C., et al. 2017, ApJ, 850, 74
Koss, M. J., Romero-Cañizales, C., Baronchelli, L., et al. 2015, ApJ, 807, 149
Koss, M. J., Assef, R., Baloković, M., et al. 2016, ApJ, 825, 85
Krolik, J. H., & Begelman, M. C. 1988, ApJ, 329, 702
Liu, Y., & Li, X. 2014, ApJ, 787, 52
—. 2015, MNRAS, 448, L53
Madejski, G., Życki, P., Done, C., et al. 2000, ApJ, 535, L87
Magdziarz, P., & Zdziarski, A. A. 1995, MNRAS, 273, 837
Malizia, A., Landi, R., Molina, M., et al. 2016, MNRAS, 460, 19
Markwardt, C. B., Tueller, J., Skinner, G. K., et al. 2005, ApJ, 633, L77
Masini, A., Comastri, A., Puccetti, S., et al. 2017, A&A, 597, A100
Matt, G., Bianchi, S., D'Ammando, F., & Martocchia, A. 2004, A&A, 421, 473
Mitsuda, K., Bautz, M., Inoue, H., et al. 2007, PASJ, 59, 1
Murphy, K. D., & Yaqoob, T. 2009, MNRAS, 397, 1549

Nenkova, M., Sirocky, M. M., Ivezić, Ž., & Elitzur, M. 2008a, ApJ, 685, 147
Nenkova, M., Sirocky, M. M., Nikutta, R., Ivezić, Ž., & Elitzur, M. 2008b, ApJ, 685, 160
Oda, S., Tanimoto, A., Ueda, Y., et al. 2017, ApJ, 835, 179
Piconcelli, E., Bianchi, S., Guainazzi, M., Fiore, F., & Chiaberge, M. 2007, A&A, 466, 855
Puccetti, S., Comastri, A., Fiore, F., et al. 2014, ApJ, 793, 26
Ricci, C., Ueda, Y., Koss, M. J., et al. 2015, ApJ, 815, L13
Ricci, C., Ueda, Y., Paltani, S., et al. 2014, MNRAS, 441, 3622
Ricci, C., Bauer, F. E., Treister, E., et al. 2017, MNRAS, 468, 1273
Rivers, E., Baloković, M., Arévalo, P., et al. 2015, ApJ, 815, 55
Schurch, N. J., Roberts, T. P., & Warwick, R. S. 2002, MNRAS, 335, 241
Severgnini, P., Caccianiga, A., Della Ceca, R., et al. 2011, A&A, 525, A38
Smith, R. K., Brickhouse, N. S., Liedahl, D. A., & Raymond, J. C. 2001, ApJ, 556, L91
Stalevski, M., Ricci, C., Ueda, Y., et al. 2016, MNRAS, 458, 2288
Tanimoto, A., Ueda, Y., Kawamuro, T., & Ricci, C. 2016, PASJ, 68, S26
Tazaki, F., Ueda, Y., Terashima, Y., & Mushotzky, R. F. 2011, ApJ, 738, 70
Tazaki, F., Ueda, Y., Terashima, Y., Mushotzky, R. F., & Tombesi, F. 2013, ApJ, 772, 38
Treister, E., Urry, C. M., & Virani, S. 2009, ApJ, 696, 110
Tueller, J., Mushotzky, R. F., Barthelmy, S., et al. 2008, ApJ, 681, 113
Tueller, J., Baumgartner, W. H., Markwardt, C. B., et al. 2010, ApJS, 186, 378
Turner, T. J., George, I. M., Nandra, K., & Mushotzky, R. F. 1998, ApJ, 493, 91
Turner, T. J., & Miller, L. 2009, A&A Rev., 17, 47
Ueda, Y. 2015, Proceeding of the Japan Academy, Series B, 91, 175
Ueda, Y., Akiyama, M., Hasinger, G., Miyaji, T., & Watson, M. G. 2014, ApJ, 786, 104
Ueda, Y., Akiyama, M., Ohta, K., & Miyaji, T. 2003, ApJ, 598, 886
Ueda, Y., Eguchi, S., Terashima, Y., et al. 2007, ApJ, 664, L79
van den Bosch, R. C. E. 2016, ApJ, 831, 134
Vasudevan, R. V., Brandt, W. N., Mushotzky, R. F., et al. 2013, ApJ, 763, 111
Vasudevan, R. V., & Fabian, A. C. 2009, MNRAS, 392, 1124
Verner, D. A., & Ferland, G. J. 1996, ApJS, 103, 467
Verner, D. A., Ferland, G. J., Korista, K. T., & Yakovlev, D. G. 1996, ApJ, 465, 487
Winter, L. M., Mushotzky, R. F., Reynolds, C. S., & Tueller, J. 2009a, ApJ, 690, 1322
Winter, L. M., Mushotzky, R. F., Terashima, Y., & Ueda, Y. 2009b, ApJ, 701, 1644
Winter, L. M., Mushotzky, R. F., Tueller, J., & Markwardt, C. 2008, ApJ, 674, 686
Wright, E. L., Eisenhardt, P. R. M., Mainzer, A. K., et al. 2010, AJ, 140, 1868
Yang, Y., Wilson, A. S., Matt, G., Terashima, Y., & Greenhill, L. J. 2009, ApJ, 691, 131
Yaqoob, T. 2012, MNRAS, 423, 3360
Yaqoob, T., Tatum, M. M., Scholtes, A., Gottlieb, A., & Turner, T. J. 2015, MNRAS, 454, 973